\documentclass[twocolumn,times]{aastex62}

\usepackage{CJKutf8}

\usepackage{amsmath}
\usepackage{amssymb}
\usepackage{amsfonts}
\usepackage{graphicx}
\usepackage{wasysym}
\usepackage{mdframed}
\usepackage[T1]{fontenc}

\hypersetup{linkcolor=cyan,citecolor=blue,filecolor=cyan,urlcolor=blue}

\newcommand{\change}[1]{\textrm{#1}}

\newcommand{\meth}{\mbox{CH$_3$OH}}

\newcommand{\fmh}{\mbox{H$_2$CO}}

\newcommand{\water}{\mbox{H$_2$O}}

\newcommand{\kms}{\mbox{km\,s$^{-1}$}}

\newcommand{\sqc}{\mbox{cm$^{-2}$}}
\newcommand{\cc}{\mbox{cm$^{-3}$}}

\newcommand{\msol}{\mbox{$M_\odot$}}
\newcommand{\msolpyr}{\mbox{$M_\odot$\,yr$^{-1}$}}
\newcommand{\mjypbm}{\mbox{mJy\,beam$^{-1}$}}
\newcommand{\ujypbm}{\mbox{$\mu$Jy\,beam$^{-1}$}}
\newcommand{\jypbm}{\mbox{Jy\,beam$^{-1}$}}
\newcommand{\hii}{\mbox{H\,{\sc ii}}}
\newcommand{\vlsr}{\mbox{$V_\text{lsr}$}}
\newcommand{\clt}{\mbox{class\,{\sc ii}}}

\newcommand{\Clt}{\mbox{Class\,{\sc ii}}}

\newcommand{\ctw}{the 20~\kms{} cloud}
\newcommand{\cfi}{the 50~\kms{} cloud}
\newcommand{\Ctw}{The 20~\kms{} cloud}
\newcommand{\Cfi}{The 50~\kms{} cloud}

\newcommand{\gzp}{G0.253+0.016}

\received{- -, --}
\revised{- -, --}
\accepted{- -, --}
\submitjournal{ApJ}

\shorttitle{High-Mass Star Formation in Central Molecular Zone}
\shortauthors{Lu et al.}

\begin{document}
\begin{CJK}{UTF8}{gbsn}

\title{A Census of Early Phase High-Mass Star Formation in the Central Molecular Zone}

\correspondingauthor{Xing Lu}
\email{xinglv.nju@gmail.com, xing.lu@nao.ac.jp}

\author[0000-0003-2619-9305]{Xing Lu (吕行)}
\affiliation{National Astronomical Observatory of Japan, 2-21-1 Osawa, Mitaka, Tokyo, 181-8588, Japan}

\author{Elisabeth A.\ C.\ Mills}
\affiliation{Physics Department, Brandeis University, 415 South Street, Waltham, MA 02453, USA}

\author{Adam Ginsburg}
\affiliation{National Radio Astronomy Observatory, 1003 Lopezville Rd., Socorro, NM 87801, USA}

\author{Daniel L.\ Walker}
\affiliation{Joint ALMA Observatory, Alonso de C\'{o}rdova 3107, Vitacura 763 0355, Santiago, Chile}
\affiliation{National Astronomical Observatory of Japan, 2-21-1 Osawa, Mitaka, Tokyo, 181-8588, Japan}

\author{Ashley T.~Barnes}
\affiliation{Argelander-Institut f\"ur Astronomie, Universit\"at Bonn, Auf dem H\"ugel 71, D-53121 Bonn, Germany}

\author{Natalie Butterfield}
\affiliation{Green Bank Observatory, 155 Observatory Rd, P.O. Box 2, Green Bank, WV 24944, USA}

\author{Jonathan D. Henshaw}
\affiliation{Max-Planck-Institut f\"ur Astronomie, K\"onigstuhl 17, D-69117 Heidelberg, Germany}

\author{Cara Battersby}
\affiliation{University of Connecticut, Department of Physics, 2152 Hillside Road, Storrs, CT 06269, USA}

\author{J.\ M.\ Diederik Kruijssen}
\affiliation{Astronomisches Rechen-Institut, Zentrum f\"ur Astronomie der Universit\"at Heidelberg, M\"onchhofstra{\ss}e 12-14, D-69120 Heidelberg, Germany}

\author{Steven N.\ Longmore}
\affiliation{Astrophysics Research Institute, Liverpool John Moores University, 146 Brownlow Hill, Liverpool L3 5RF, UK}

\author{Qizhou Zhang}
\affiliation{Center for Astrophysics | Harvard \& Smithsonian, 60 Garden Street, Cambridge, MA 02138, USA}

\author{John Bally}
\affiliation{Department of Astrophysical and Planetary Sciences, University of Colorado, 389 UCB, Boulder, CO 80309, USA}

\author{Jens Kauffmann}
\affiliation{Haystack Observatory, Massachusetts Institute of Technology, Westford, MA 01886, USA}

\author{J\"urgen Ott}
\affiliation{National Radio Astronomy Observatory, 1003 Lopezville Rd., Socorro, NM 87801, USA}

\author{Matthew Rickert}
\affiliation{CIERA and Department of Physics and Astronomy, Northwestern University, 2145 Sheridan Rd, Evanston, IL 60208-3112, USA}
\affiliation{National Radio Astronomy Observatory, 1003 Lopezville Rd., Socorro, NM 87801, USA}

\author{Ke Wang}
\affiliation{Kavli Institute for Astronomy and Astrophysics, Peking University, 5 Yiheyuan Road, Haidian District, Beijing 100871, P.\ R.\ China}

\begin{abstract} % =<250 words
We present new observations of $C$-band continuum emission and masers to assess high-mass ($>$8~\msol{}) star formation at early evolutionary phases in the inner 200~pc of the Central Molecular Zone (CMZ) of the Galaxy. The continuum observation is complete to free-free emission from stars above 10--11~\msol{} in 91\% of the covered area. We identify 104 compact sources in the continuum emission, among which five are confirmed ultra-compact \hii{} regions, 12 are candidates of ultra-compact \hii{} regions, and the remaining 87 sources are mostly massive stars in clusters, field stars, evolved stars, pulsars, extragalactic sources, or of unknown nature that is to be investigated. We detect \clt{} \meth{} masers at 23 positions, among which six are new detections. We confirm six known \fmh{} masers in two high-mass star forming regions, and detect two new \fmh{} masers toward the Sgr~C cloud, making it the ninth region in the Galaxy that contains masers of this type. In spite of these detections, we find that current high-mass star formation in the inner CMZ is only taking place in seven isolated clouds. The results suggest that star formation at early evolutionary phases in the CMZ is about 10 times less efficient than expected by the dense gas star formation relation, which is in line with previous studies that focus on more evolved phases of star formation. This means that if there will be any impending, next burst of star formation in the CMZ, it has not yet begun.
\end{abstract}


%%%%%%%%%%%%%%%%%%%%%
\section{INTRODUCTION}\label{sec:intro}
Observations toward the Central Molecular Zone (CMZ), the inner $\sim$500~pc of the Galaxy, suggest a large amount of molecular gas \citep[$>$10$^7$~\msol{}, mean density $\sim$10$^4$~\cc{};][]{bally1987,longmore2013a}. However, the measured star formation rate (SFR) in the CMZ is about 10 times lower than expected by the dense gas star formation relation extrapolated from the nearby molecular clouds \citep{yusefzadeh2009,an2011,immer2012a,longmore2013a,barnes2017}. Various mechanisms (or combinations of them) have been suggested to explain the inefficient star formation in the CMZ, including inhibition of gas collapse by strong turbulence \citep{kruijssen2014,dale2019}, episodic star formation regulated by Galactic dynamics \citep{kruijssen2014,KK2015,krumholz2017,meidt2018,kruijssen2019}, and higher density thresholds for star formation \citep{kruijssen2014,federrath2016,krumholz2017}.

Despite the advances in theoretical models, star formation in the CMZ is not well characterized observationally.  Previous studies have used infrared luminosities or young stellar objects (YSOs) in infrared bands of the CMZ to measure star formation \citep[e.g.,][]{yusefzadeh2009,an2011,immer2012a,barnes2017}. However, these approaches suffer from heavy extinction in the infrared bands toward the Galactic Center \citep{barnes2017} and contamination from more evolved stellar populations \citep{koepferl2015}. More recently, star formation in a few clouds in the CMZ was characterized at high angular resolution by using masers and ultra-compact (UC) \hii{} regions \citep{lu2015b,lu2019a,kauffmann2017a}, which are free of extinction and trace early phase high-mass ($>$8~\msol{}) star formation. A few CMZ-wide surveys of masers have also provided important information on the distribution of star formation in the CMZ \citep[e.g.,][]{caswell2010,chambers2014,cotton2016,rickert2019}.

Here we report high angular resolution, high sensitivity observations of $C$-band line and continuum emission toward the inner CMZ carried out with the NRAO\footnote{The National Radio Astronomy Observatory is a facility of the National Science Foundation operated under cooperative agreement by Associated Universities, Inc.} Karl G.\ Jansky Very Large Array (VLA). Our observations feature a large surveyed area combined with high resolution, which enables a comprehensive census of masers and UC \hii{} regions. We aim to study high-mass star formation at early evolutionary phases, in which protostars are deeply embedded in molecular gas and dust, when near to mid-infrared emission from star formation is weak due to absorption. Therefore they are best traced by masers and free-free emission from embedded UC \hii{} regions. In \autoref{sec:obs}, we introduce our observations. In \autoref{sec:results}, we report maser and continuum source detections, and specifically, the detection of several new masers toward high-mass star forming regions. Then in \autoref{sec:discussion} we identify candidates of UC \hii{} regions from the continuum emission, and discuss the implications for star formation in the CMZ. We conclude and summarize our findings in \autoref{sec:conclusions}. Throughout this paper we adopt a distance of 8.1~kpc to the CMZ \citep{gravity2018}.

\begin{deluxetable*}{ccccc}
%\tabletypesize{\scriptsize}
\tablecaption{Summary of VLA Observations.\label{tab:obs}}
\tablewidth{0pt}
\tablehead{
\colhead{Obs.~Date} & Field Indices\tablenotemark{a} & No.\ of Unflagged Antennas & $uv$-distance ($k\lambda$) & Calibrators\tablenotemark{b}
}
\startdata
Aug 21 2016 & 8--17,19,21  & 23 & 2--230 & 3C286, J1744$-$3116 \\
Sep 12 2016 & 18,20,22--31 & 23 & 2--253 & 3C286, J1744$-$3116 \\
Oct 04 2016 & 1--7         & 25 & 4--472 & 3C286, J1744$-$3116
\enddata
\tablenotetext{a}{Field indices are marked in \autoref{fig:overview}.}
\tablenotetext{b}{Bandpass/flux calibrator and phase calibrator, respectively.}
\end{deluxetable*}

\begin{figure*}[!t]
\centering
\includegraphics[width=1.0\textwidth]{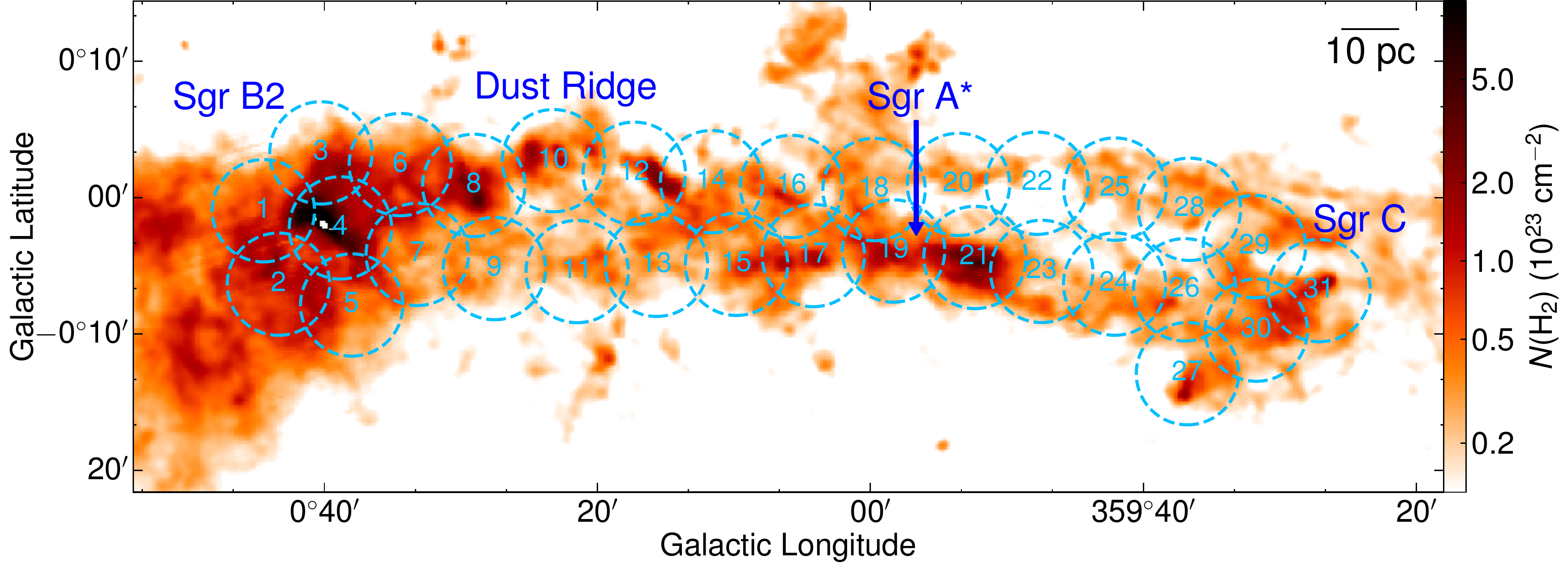}
\caption{Spatial coverage of the VLA observations. Cyan dashed circles show the fields, with field indices marked at the circle centers. The position of Sgr~A* is marked by an arrow, while approximate locations of the other three tiles (Sgr~B2, Dust Ridge, Sgr C) are labeled. The background image shows molecular gas column densities derived from \textit{Herschel} (\citealt{battersby2011}).}
\label{fig:overview}
\end{figure*}

%%%%%%%%%%%%%%%%%%%%%
\section{OBSERVATIONS AND DATA REDUCTION}\label{sec:obs}

The VLA observations were carried out in the B configuration in August--October 2016 with the project code 16A-173. A brief summary of the observations is shown in \autoref{tab:obs}. The $C$-band receiver was used to cover six lines, including two \meth{} lines at the rest frequencies of 6.668~GHz and 5.005~GHz, a radio recombination line H(111)$\alpha$ at the rest frequency of 4.744~GHz, and three formaldehyde isotopologue lines \fmh{}, H$_2$$^{13}$CO, H$_2$C$^{18}$O between rest frequencies of 4.389 and 4.830~GHz. For each line, a velocity range of $\pm$200~\kms{} was covered. In addition to the lines, a total of 16 wide-band spectral windows was used to cover 2~GHz wide continuum between 4.2 and 6.9~GHz, but owing to strong radio frequency interference (RFI), the available bandwidth for continuum imaging is less, with a typical value of 1.7~GHz.

We targeted the inner 200~pc of the CMZ (\autoref{fig:overview}), where high-mass stars are most likely to be forming given the high gas column densities \citep{kauffmann2010}. A total of 31 fields were observed. The fields are not Nyquist sampled, as we do not target extended structures above the largest recoverable angular scale of $\sim$1\arcmin{}, therefore mosaicked imaging is not required. Among these fields, seven were used to cover the Sgr~B2 region in a hexagonal pattern (fields 1--7), and the others were used to cover the high column density ($\gtrsim$10$^{23}$~\sqc{}) regions roughly along the ring-like 100-pc structure seen in \textit{Herschel} infrared emission images \citep{molinari2010}. Each field was integrated for about 3.5 minutes in a snapshot mode. The FWHM size of the primary beam of the VLA at the central frequency of 5.56~GHz is 7\farcm{5}, and this size varies from 8$'$ to 7$'$ from the lowest to the highest observed frequencies.

We calibrated the data following standard procedures using CASA 5.4.0. Note that at this frequency band, significant RFI exists\footnote{\url{https://science.nrao.edu/facilities/vla/docs/manuals/obsguide/rfi}}. Therefore, we first used the \textit{rflag} algorithm implemented in CASA to automatically identify and flag RFI in the calibrators, then performed a manual flagging to remove any other significant RFI.

Then we imaged the lines and continuum using CASA 5.4.0. For the primary target line, the \meth{} line at 6.668~GHz, we first performed Hanning smoothing to merge every two velocity channels in the calibrated data, in order to remove the effects of Gibbs ringing. The resulting channel width is 0.35~\kms{}. We used the $uvcontsub$ task in CASA to subtract the continuum. Then we imaged each field separately using the $tclean$ task, with the Briggs weighting and a robust number of 0.5. When there is a strong maser detected in the field, we performed self calibration using its peak channel, and applied the calibration tables to all the other channels. For the seven fields covering the Sgr~B2 region, two antennas were in longer baselines in the array during the observation, which result in a smaller synthesized beam. To get a larger beam size that is consistent with the other fields, we applied $uv$ tapering to the seven fields, and achieved an image rms of 8~\mjypbm{} with a beam size of 1\farcs{2}$\times$1\farcs{0} in a channel width of 0.35~\kms{}. For the other fields, we skipped $uv$ tapering, and achieved an image rms of 7--9~\mjypbm{} with a beam size of 2\farcs{1}$\times$1\farcs{2} in a channel width of 0.35~\kms{}.

Similarly, we imaged the \meth{} line at 5.005~GHz, the H(111)$\alpha$ line, and the three \fmh{} isotopologue lines for each field. Among them, the \fmh{} line is more complex: toward several positions in Sgr~B2, this line presents deep absorption that extends beyond the velocity coverage of the spectral window ($\pm$200~\kms{}). This is verified in the wide-band spectral window (120~MHz bandwidth) that covers this frequency, where absorption up to $\pm$400~\kms{} is found. Therefore, the continuum subtraction using the $uvcontsub$ task creates artificial positive intensities for local maxima (less absorption) between deep absorption features. In \autoref{subsec:results_h2co}, we will argue that this does not affect our search for \fmh{} masers, but this issue will likely hinder any further investigation of \fmh{} absorption toward these positions in Sgr~B2 using these data.

\begin{deluxetable*}{ccccc}
%\tabletypesize{\scriptsize}
\tablecaption{Comparison of CMZ Radio Continuum Surveys.\label{tab:surveys}}
\tablewidth{0pt}
\tablehead{
\colhead{\ } & Band/Frequency & Resolution & Sensitivity\tablenotemark{a}
}
\startdata
This work                      & $C$-band/5.56~GHz  & 1\arcsec{} & 0.025--0.5~\mjypbm{} \\
\citet{becker1994}     & $C$-band/4.9~GHz     & 4\arcsec{} & $\geq$2.5~\mjypbm{} \\
\citet{zoonematkermani1990} & $L$-band/1.4~GHz & 5\arcsec{} & $\geq$1--2~\mjypbm{} \\
\citet{yusefzadeh2004} & $L$-band/1.4~GHz & 10\arcsec{} & $\geq$0.16~\mjypbm{} \\
\citet{lazio2008}        & $L$-band/1.4~GHz    & 2\arcsec{}  & $\geq$0.05~\mjypbm{} \\
\citet{lang2010}          & $L$-band/1.4~GHz    & 15\arcsec{} & 3--4~\mjypbm{}
\enddata
\tablenotetext{a}{Lower limits of sensitivities are noted with `$\geq$'. Real noise levels can be at least 10 times higher.}
\end{deluxetable*}

Finally, we imaged the continuum emission. We first identified and flagged RFI using the \textit{rflag} algorithm, and then manually flagged any residual significant RFI. As a number of fields contain known extended continuum emission, we used a different imaging strategy than for the lines. We split the 31 fields into four tiles and mosaicked each one: the first tile mostly covers Sgr~B2 and includes fields 1--7; the second tile covers the Dust Ridge (and Sgr B1) and includes fields 8--15; the third tile covers Sgr~A and includes fields 16--23; and the fourth tile covers Sgr~C and includes fields 24--31. We used the \textit{tclean} task, with n-term of 2, multi-frequency synthesis, and multi-scale parameters of [0,3,10,30], and gridded the tiles in a cell size of 0\farcs{3}. The n-term of 2 was used to fit a linear function to the data over a frequency range between 4.2 and 6.9~GHz to obtain spectral indices $\alpha$ as well as the uncertainty on the fit, $\sigma(\alpha)$. The spectral index $\alpha$ is defined by $S_\nu\propto\nu^{\alpha}$, where $S_\nu$ is the specific flux at the frequency $\nu$. In addition, for the Sgr~A and Sgr~B2 tiles, where the continuum emission is sufficiently strong, we performed self calibration to improve the dynamic range, although we found that the signal-to-noise ratio is only improved by a factor of $<$2, likely because the noise is dominated by partially resolved extended structures. The thermal noise level, 20~\ujypbm{}, is not achieved in any of the continuum maps. The measured noise represented by the rms is as low as 25~\ujypbm{} toward a few small regions where the continuum emission is undetected (e.g., in the southern part of the Sgr~C tile), and as high as 500~\ujypbm{} next to Sgr~A* and Sgr~B2. In \autoref{subsec:results_compact} we will construct localized noise maps to account for the varying noise. As a final step, the Stokes-$I$ images (with a central frequency at 5.56~GHz), the spectral index images ($\alpha$), and the spectral index uncertainty images ($\sigma(\alpha)$) were corrected for the primary beam response using the \textit{widebandpbcor} task. \change{The continuum images are publicly available at\dataset[10.5281/zenodo.3361116]{https://doi.org/10.5281/zenodo.3361116}.}

In \autoref{tab:surveys} we compare our continuum observation with several radio continuum surveys that have covered the CMZ. Our observation has a higher angular resolution and better sensitivity than the $C$-band continuum survey of \citet{becker1994}, which is part of the Multi-Array Galactic Plane Imaging Survey (MAGPIS), as well as the four $L$-band surveys. The flux of optically thin free-free emission from (UC) \hii{} regions has a weak dependency on observed frequencies (i.e., a spectral index of $-$0.1 or slightly higher), therefore the sensitivities at the different frequencies can be directly compared. It then follows that among the surveys in \autoref{tab:surveys} our observations deliver the best sensitivity combined with the highest angular resolution that is optimal for the search of UC \hii{} regions.

To check the quality of bandpass and flux calibration, we imaged the continuum emission of the phase calibrator, J1744$-$3116. This quasar presents a consistent flux of 0.74~Jy across the three epochs of observations, and a measured spectral index between $-0.03$ (the first two epochs) and $0.02$ (the last epoch). If the spectral index of the quasar is invariant during the observations, this suggests a systematic uncertainty of 0.05 in the measured spectral indices.

\begin{figure*}[!t]
\centering
\includegraphics[width=0.9\textwidth]{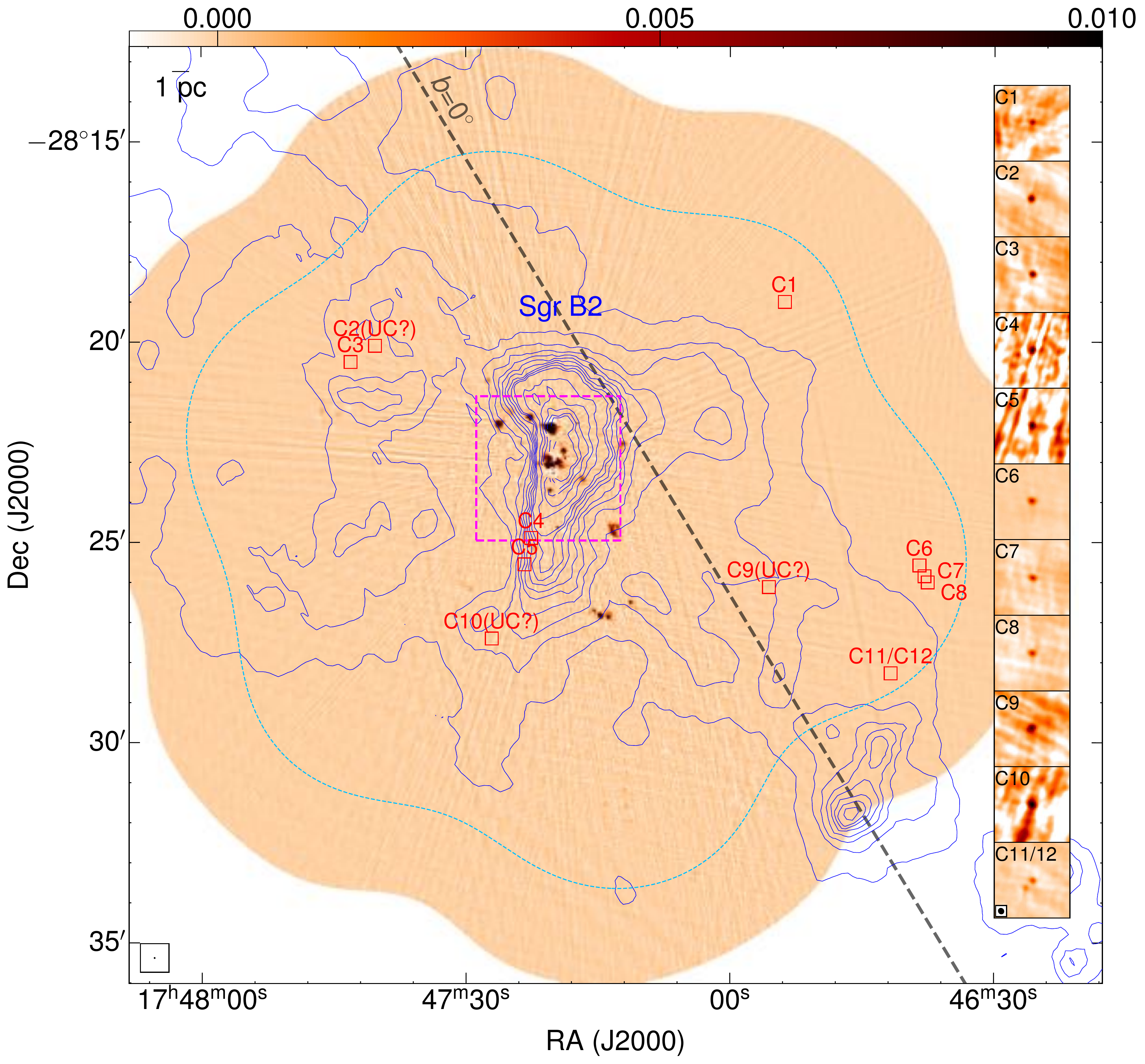}
\caption{Overview of the Sgr~B2 tile. The background image shows the VLA $C$-band continuum emission \change{in the linear scale}, which is truncated at an intensity of 0.01~\jypbm{} to highlight faint sources. \change{The unit of the color bar attached to the image top is \jypbm{}.} The cyan dashed loop shows the FWHM of primary beam response of the mosaic. Small boxes mark identified compact sources, among which UC \hii{} candidates and confirmed UC \hii{} regions are labeled with `UC?' and `UC', respectively. Their zoom-in views are shown in the insets aligned on the right. Each inset, centered on the compact source, is 15\arcsec{} across, and the color scale is adjusted to match the peak intensity of the source. In the last inset (C11/12 in this case), the synthesized beam is plotted in the bottom left corner. The large dashed box marks the region where masers are detected, and a zoom-in view is in \autoref{fig:maser}. Blue contours show column densities derived from the \textit{Herschel} data (\citealt{battersby2011}), between [0.5, 5]$\times$10$^{23}$~\sqc{} in steps of 0.5$\times$10$^{23}$~\sqc{}, and then between [5, 15]$\times$10$^{23}$~\sqc{} in steps of 2$\times$10$^{23}$~\sqc{}. Names of individual clouds (e.g., Sgr~B2 in this tile) are labeled. The dashed diagonal line marks the Galactic latitude line at 0\arcdeg{}.}
\label{fig:sgrb2}
\end{figure*}

\begin{figure*}[!t]
\centering
\includegraphics[width=0.95\textwidth]{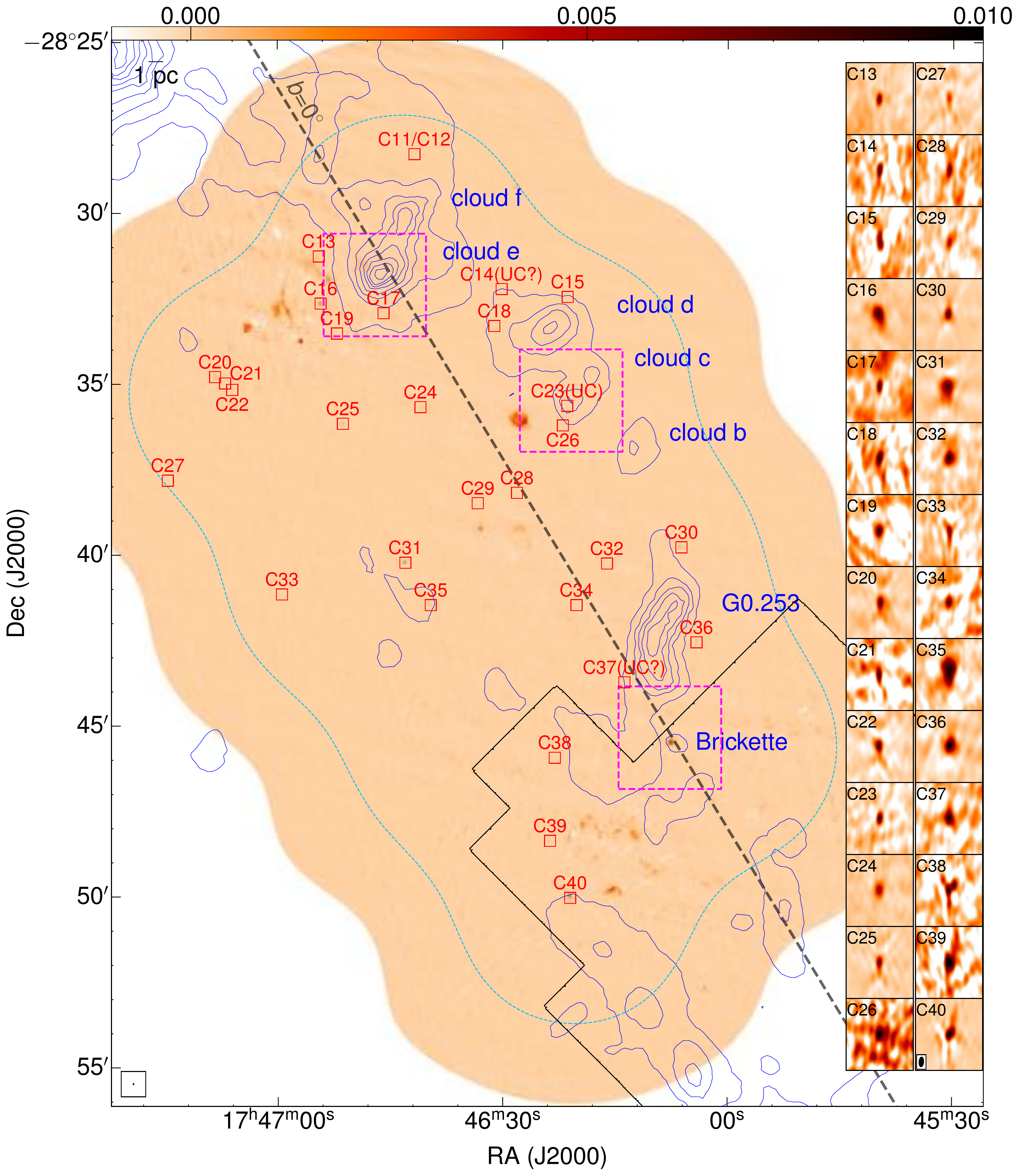}
\caption{Overview of the Dust Ridge tile. Contours and symbols are the same as in \autoref{fig:sgrb2}. The black contour shows the coverage of the HST
Paschen-$\alpha$ survey of \citet{wang2010} (see \autoref{subsubsec:disc_others}).}
\label{fig:dustridge}
\end{figure*}

\begin{figure*}[!t]
\centering
\includegraphics[width=0.95\textwidth]{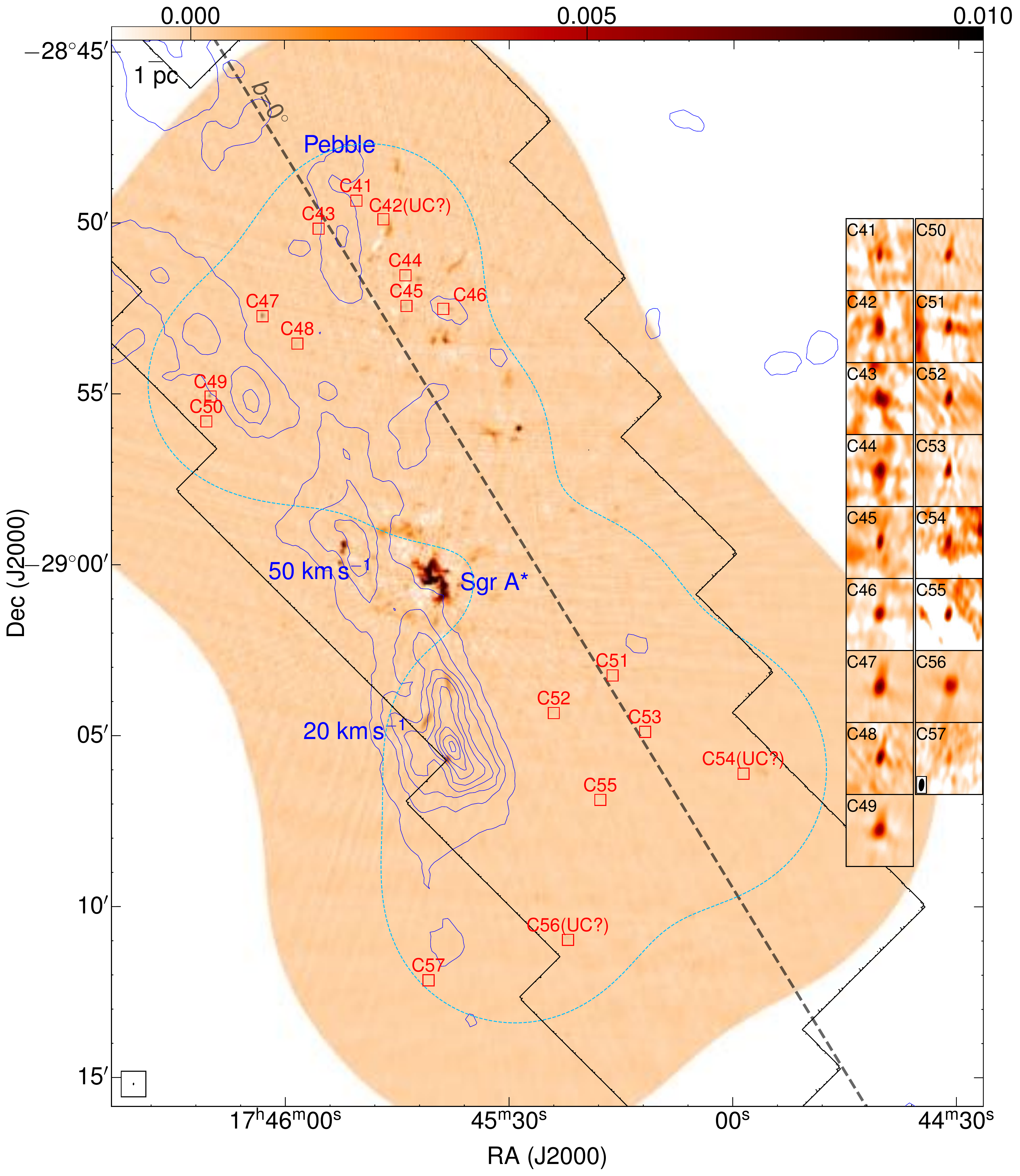}
\caption{Overview of the Sgr~A tile. Contours and symbols are the same as in \autoref{fig:sgrb2}. The black contour shows the coverage of the HST
Paschen-$\alpha$ survey of \citet{wang2010} (see \autoref{subsubsec:disc_others}). Note that the primary beam response of the mosaic, whose FWHM is represented by the green dashed loop, is unusually low around Sgr~A*. This is because the visibility data of this field is down-weighted for the imaging due to higher noise than other fields in the mosaic.}
\label{fig:sgra}
\end{figure*}

\begin{figure*}[!t]
\centering
\includegraphics[width=0.95\textwidth]{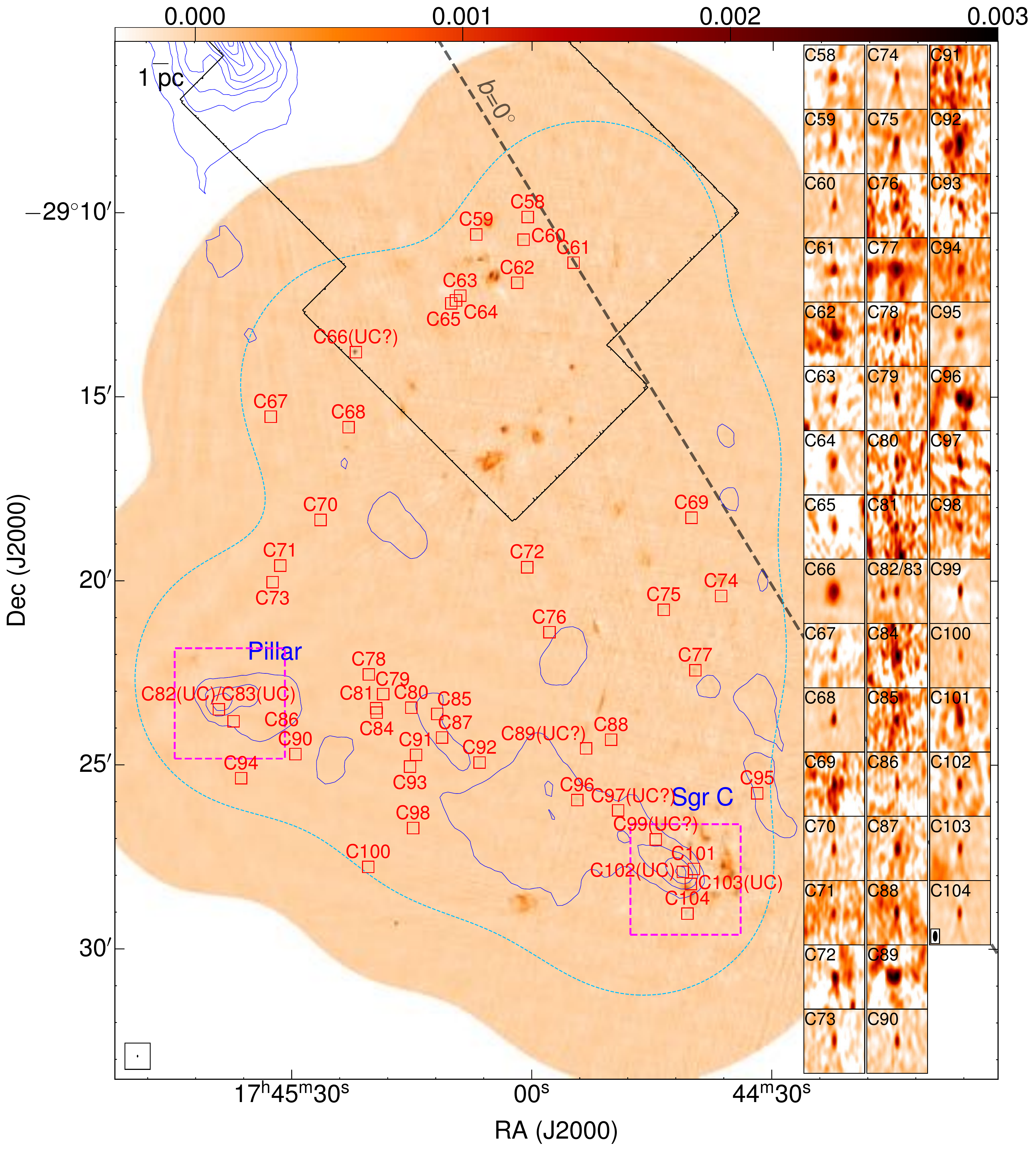}
\caption{Overview of the Sgr~C tile. Contours and symbols are the same as in \autoref{fig:sgrb2}. The black contour shows the coverage of the HST
Paschen-$\alpha$ survey of \citet{wang2010} (see \autoref{subsubsec:disc_others}).}
\label{fig:sgrc}
\end{figure*}

%%%%%%%%%%%%%%%%%%%%%%%%%
\section{RESULTS}\label{sec:results}

In this paper, we focus on potential star formation tracers, including the continuum emission that may arise from \hii{} regions, the \clt{} \meth{} maser at 6.668~GHz, and the \fmh{} maser at 4.830~GHz. The \meth{}/\fmh{}/H$_2$$^{13}$CO absorption, and non-thermal continuum emission will be discussed in forthcoming papers. The \meth{} line at 5.005~GHz, the H(111)$\alpha$ line, and the H$_2$C$^{18}$O line are not detected at the 3$\sigma$ level.

%%%%%%%%%%%%%%%
\subsection{$C$-band Continuum Emission\label{subsec:results_cont}}
The $C$-band continuum emission images are presented in Figures~\ref{fig:sgrb2}--\ref{fig:sgrc}. Bright continuum emission is detected inside the Sgr~A (\autoref{fig:sgra}) and Sgr~B2 (\autoref{fig:sgrb2}) tiles, which results in dynamic range limited imaging and significant noise. Nevertheless, we carefully compared with previous observations, and found that the well known features in these two regions are clearly recovered in our images: the mini-spiral arms around Sgr~A* \citep[e.g.,][]{lo1983,ekers1983,roberts1993,zhao2009,zhao2013,zhao2016,tsuboi2016,tsuboi2017}; the four \hii{} regions in the Sgr~A East region \citep[associated with \cfi{}; e.g.,][]{yusefzadeh2010,mills2011}; and the prominent star forming sites Sgr~B2(N), Sgr~B2(M), and Sgr~B2(S) \citep[e.g.,][]{mehringer1993,mehringer1995,gaume1995,depree1995,depree1996,depree2015}.

The weaker continuum emission in the Dust Ridge and Sgr~C tiles also morphologically agrees with previous detections, although for some sources we do not find observations in the literature at as high an angular resolution as ours. We compared with radio continuum images toward the Dust Ridge clouds \citep[e.g., G0.253$-$0.025, Dust Ridge clouds c/e/f;][]{immer2012b,rodriguez2013,mills2015,ludovici2016,butterfield2018,lu2019a} and the Sgr~C cloud \citep{forster2000,lu2019a}, and found counterparts in our $C$-band continuum images.

Our observations recover emission at angular scales of $\sim$1--60\arcsec{}, and are not sensitive to emission above the angular scale of 60\arcsec{} (2.4~pc at the distance of the CMZ). Consequently, spatially extended structures such as diffuse \hii{} regions, supernova remnants, and non-thermal filaments tend to be resolved out. This is evident toward the radio bright zone around Sgr~A* \citep[$\sim$3\arcmin{} across; specifically, the Sgr~A East supernova remnant shell;][]{zhao2016}, as well as the radio filaments projected throughout the CMZ \citep[usually a few arcmin;][]{yusefzadeh2004,lang2010}, which are mostly resolved out in our observations.

%%%%%%%%%%%%%%%
\subsection{Identification of Compact Continuum Sources}\label{subsec:results_compact}
We identified compact sources from the $C$-band continuum emission, from which we search for UC~\hii{} regions later (see \autoref{subsec:disc_uchii}). The typical size of UC \hii{} regions is $\lesssim$0.1~pc \citep{churchwell2002}, equivalent to $\lesssim$2\farcs{5}, and therefore we should only consider compact (point-like) sources in our observations of $\sim$1\arcsec{}--2\arcsec{} resolution. There are more extended ($>$2\farcs{5}) sources that correspond to known \hii{} regions \citep[e.g., in Sgr~B2(M), Sgr~B2(N), and Sgr~A East;][]{gaume1995,depree2015,mills2011}, but we excluded them in the following discussion as they represent more evolved stages of star formation.

As discussed in \autoref{sec:obs}, the continuum images are dynamic range limited, especially in the Sgr~A and Sgr~B2 tiles, where the thermal noise level ($\sim$20~\ujypbm{}) cannot be achieved. In addition, partially resolved-out sources cannot be completely cleaned and therefore increase the local rms. As a result, the rms of the continuum images varies greatly across the maps. To account for this varying rms level, we first constructed a noise map for each tile with primary-beam-corrected images, using the SExtractor package \citep{bertin1996}. As recommended in \citet{hales2012}, we used a mesh size of 27$\times$27 pixels, corresponding to $\sim$80 independent beams in one mesh given that one beam encompasses about 3 pixels in one dimension. The resulting noise map achieves a balance between reflecting local rms variations and having a statistically robust sample of independent measurements within each mesh. The median noise level within the FWHM is 229~\ujypbm{} for the Sgr~B2 tile, 105~\ujypbm{} for the Dust Ridge tile, 170~\ujypbm{} for the Sgr~A tile, and 53~\ujypbm{} for the Sgr~C tile.

To identify compact sources, we employed the BLOBCAT software \citep{hales2012}, which has been adopted in several radio continuum surveys \citep[e.g.,][]{bihr2016,wang2018}. BLOBCAT utilizes the flood fill algorithm to detect and catalogue blobs, or islands of pixels representing sources, in 2D images. The noise map and the primary-beam-corrected image for each tile were fed into BLOBCAT. We set the detection threshold to 5$\sigma$, and the flooding threshold to 2.6$\sigma$ as recommend in \citet{hales2012}. In addition, to search for compact sources, we put an upper limit of 800 pixels for the source area, which corresponds to a radius of $\sim$0.2~pc. This upper limit is slightly relaxed compared with the usually adopted size of UC~\hii{} regions ($\lesssim$0.1~pc) to allow for some sources that appear spatially blended due to strong side-lobes. The largest radius we actually found is 0.16~pc (for C56, see \autoref{tab:uchii}).

We further performed a visual inspection of the identified sources, and rejected artifacts including apparent side-lobes, partially resolved-out extended sources, and irregular shaped sources that are not point-like. Sources outside of the FWHM of the primary beam response were also excluded. Nonetheless, toward extended structures that are significantly resolved out, residual emission may still be misclassified as compact sources. For example, C42 spatially overlaps with the Arched filaments \citep{yusefzadeh1984,lang2001} that are mostly resolved-out in our data, therefore could just be the residual.

Finally, we identified 104 compact sources. We took peak coordinates, peak intensities, and integrated fluxes from the output of BLOBCAT. We calculated areas of the sources using the number of pixels reported by BLOBCAT, \change{subtracted the beam area quadratically (assuming a Gaussian beam with major and minor axes reported in \autoref{sec:obs})}, and obtained beam-deconvolved effective radii of the compact sources. Spectral indices and their uncertainties at the peak intensities, as derived using the multi-term \textit{tclean} of the $C$-band continuum in \autoref{sec:obs}, were extracted from the data products. These properties are listed in \autoref{tab:uchii}. The positions of the compact sources are marked by red boxes in Figures~\ref{fig:sgrb2}--\ref{fig:sgrc}, where some closely packed ones are marked by one single box.

Note that we have avoided the Sgr~B2 region (the dashed box in \autoref{fig:sgrb2}), whose UC (and hyper-compact, HC) \hii{} population has been studied with higher angular resolution and better sensitivity than ours as well as supplemental evidence from radio recombination lines \citep[e.g.,][]{gaume1995,depree1996,sewilo2004,zhao2011,depree2015}. A total of 41 UC/HC \hii{} regions are previously identified in Sgr~B2 \citep{gaume1995,depree2015}.

Similarly, we have avoided the Sgr~A* region (the central brightest part in \autoref{fig:sgra}), where our $C$-band continuum image is dynamic range limited and spatial filtering of the VLA causes significant artifacts. In addition, due to the high noise level in the visibility data, this region is significantly down-weighted among the mosaicked pointings in \autoref{fig:sgra}, and lies mostly outside of the FWHM of the primary beam response. Although we recover a few known compact structures \citep[e.g., the four \hii{} regions in Sgr~A East, the compact radio source G$-$0.04$-$0.12;][]{mills2011}, we do not discuss them further.

\change{As shown in Figures~\ref{fig:sgrb2}--\ref{fig:sgrc}, the identified continuum sources are not randomly distributed among the four tiles as well as within the tiles. For example, there is a concentration of sources, from C78 to C98, toward the lower left region of the Sgr~C tile in \autoref{fig:sgrc}. This is because we use the detection threshold of 5$\sigma$ where the $\sigma$ value is position-dependent. In this case, more sources tend to be identified where the noise level is lower. If we use a fixed detection threshold of five times the median noise level of the Sgr B2 tile (the highest noise among the four tiles), the numbers of identified sources in the four tiles are 11, 12, 13, and 8, respectively, which are within the uncertainty of $\sim$3 around the mean value of 11 assuming Poisson statistics. Therefore, the non-uniformity among the tiles, if there is any, is not clear based on the current data. The lower left region of the Sgr~C tile has the lowest noise among all the images ($\sim$30~\ujypbm{}), therefore we were able to identify a large number of sources.}

Assuming the continuum emission is solely contributed by optically thin free-free emission from UC~\hii{} regions, we estimated the detection limit in terms of stellar masses powering the UC~\hii{} regions. With a characteristic election temperature of 8000~K, and following \citet{mezger1974}, we calculated ionizing photon rates corresponding to the 5$\sigma$ detection threshold, in which we used the median rms level in each tile as the 1$\sigma$ level. Then by comparing with the expected ionizing photon rates of ZAMS stars \citep[e.g.,][]{davies2011}, we obtained the detection limit of the continuum emission, which ranges from 10.1~\msol{} in the Sgr~C tile (where the median noise level is the lowest) to 11.6~\msol{} in the Sgr~B2 tile (where the median noise level is the highest). Around Sgr~A* and Sgr~B2 where the rms is higher than 320~\ujypbm{}, the 5$\sigma$ detection limit is $\gtrsim$12~\msol{}, though such regions are a small fraction of the coverage area ($<$9\%) and we have avoided these two regions in our identification. Therefore, except for small regions around bright continuum sources in the Sgr~B2 and Sgr~A tiles, the continuum observation is complete to free-free emission from all stars above 10--11~\msol{} (B1 type and earlier) in 91\% of the covered area.

Several compact sources have been reported in previous radio interferometer observations (not necessarily using the same frequency band as ours; e.g., \citealt{yusefzadeh1987a}, \citealt{forster2000}, \citealt{lang2001}, \citealt{lazio2008}, \citealt{rodriguez2013}, \citealt{immer2012b}, \citealt{lu2019a}). We listed corresponding references and identifiers in \autoref{tab:uchii}. In total, 24 out of the 104 compact sources have been identified before, and the remaining 80 compact sources are likely new detections. Among these previous observations, \citet{lazio2008} conducted a high angular resolution ($\sim$2\arcsec{}), high sensitivity (thermal noise level $\sim$50~\ujypbm{}, although measured noise level can be 10 times higher) survey of compact radio sources at 1.4~GHz, covering a larger area than ours. 60 of their identified compact sources fall within our observation coverage, and 13 have counterparts in our catalog (see the 2LC objects in \autoref{tab:uchii}). Excluding 18 sources that are found in the Sgr~B2 and Sgr~A* regions where we have avoided, there are 29 sources that are not included in our catalog. We compared these sources with our data, and noted that most of them are diffuse in our images, and therefore were not identified as compact sources by us. This accounts for 22 of the \citet{lazio2008} sources not in our catalog. The other 7 sources not in our catalog all present steep spectra with spectral indices of $\lesssim$$-2$ \citep[Table 8 of][]{lazio2008}, therefore are too weak at $C$-band and probably missed by our observations.

%%%%%%%%%%%%%%%%%%%%%%%%%

\subsection{\meth{} Masers}\label{subsec:results_ch3oh}
\begin{figure*}[!t]
\centering
\begin{tabular}{p{0.5\textwidth}p{0.5\textwidth}}
\includegraphics[width=0.5\textwidth]{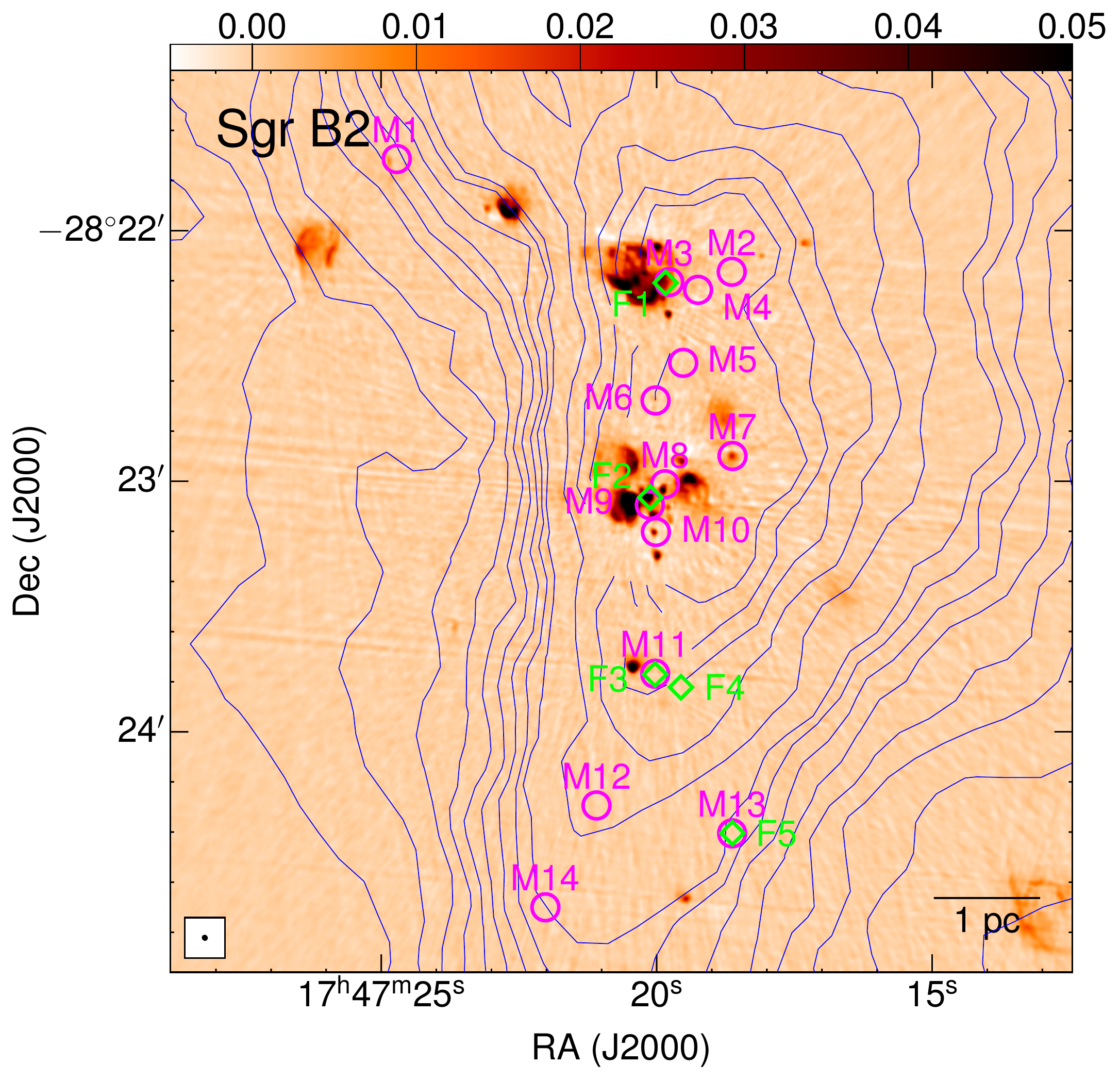} & \includegraphics[width=0.4\textwidth]{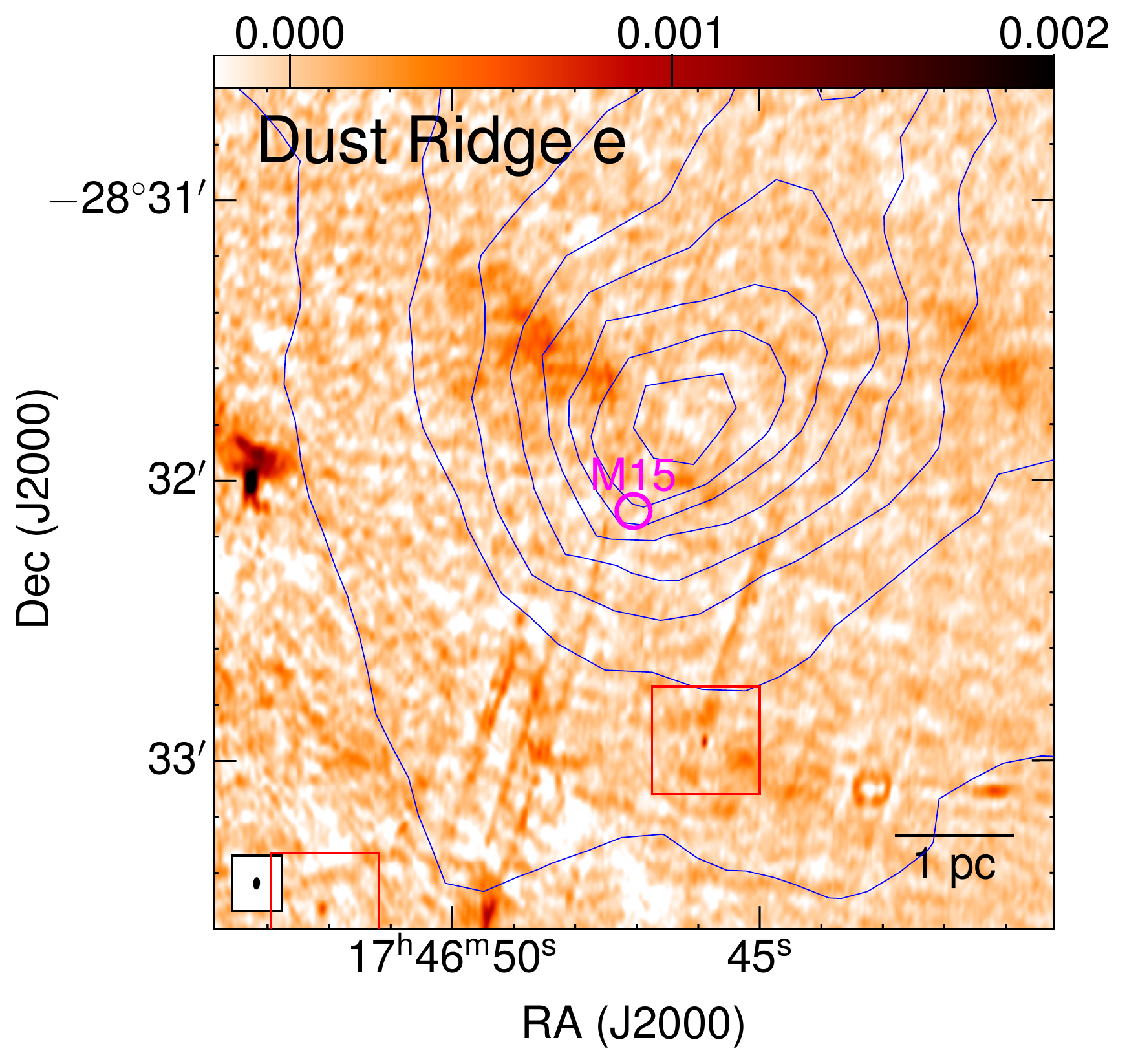} \\
\includegraphics[width=0.4\textwidth]{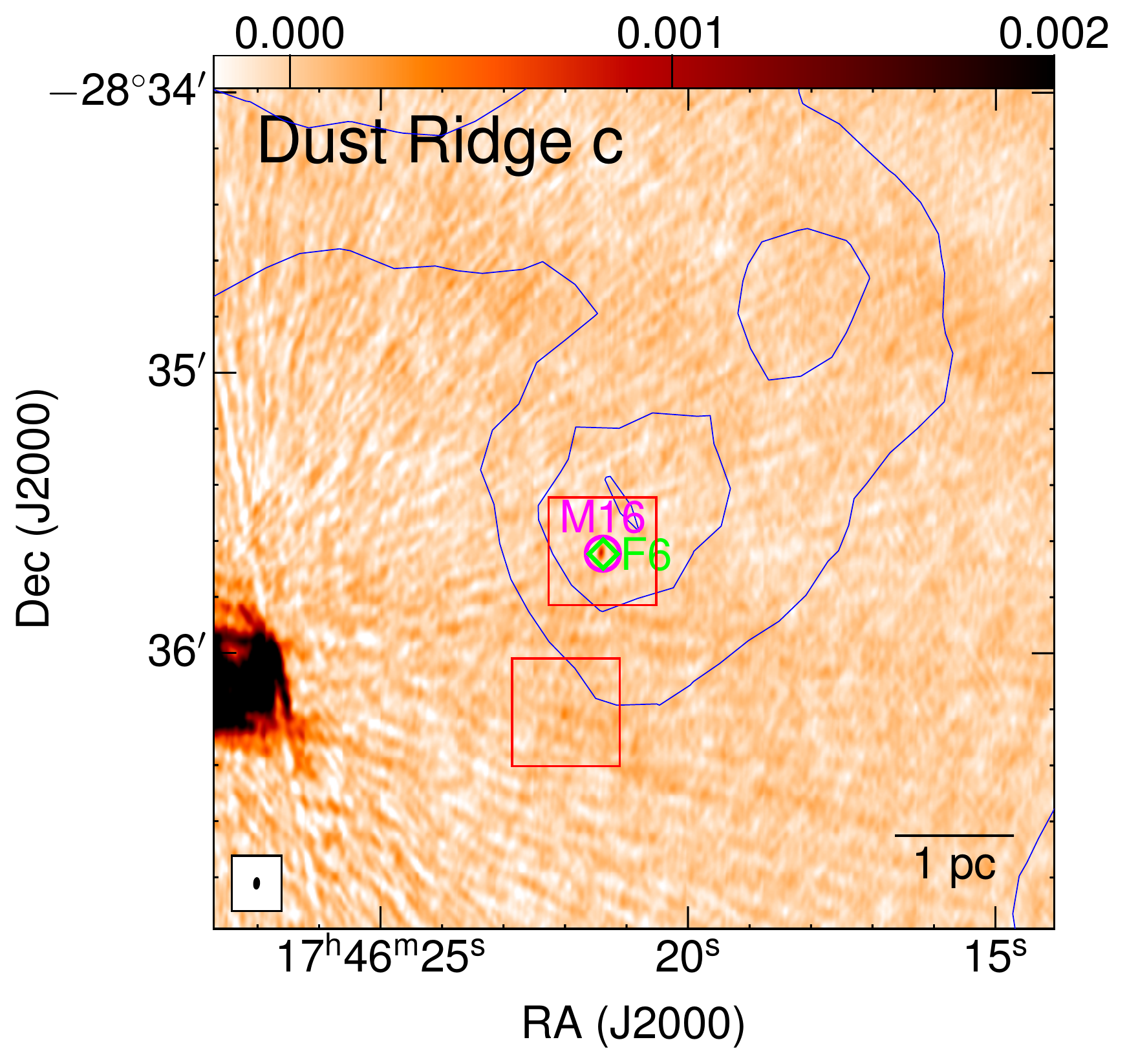} & \includegraphics[width=0.4\textwidth]{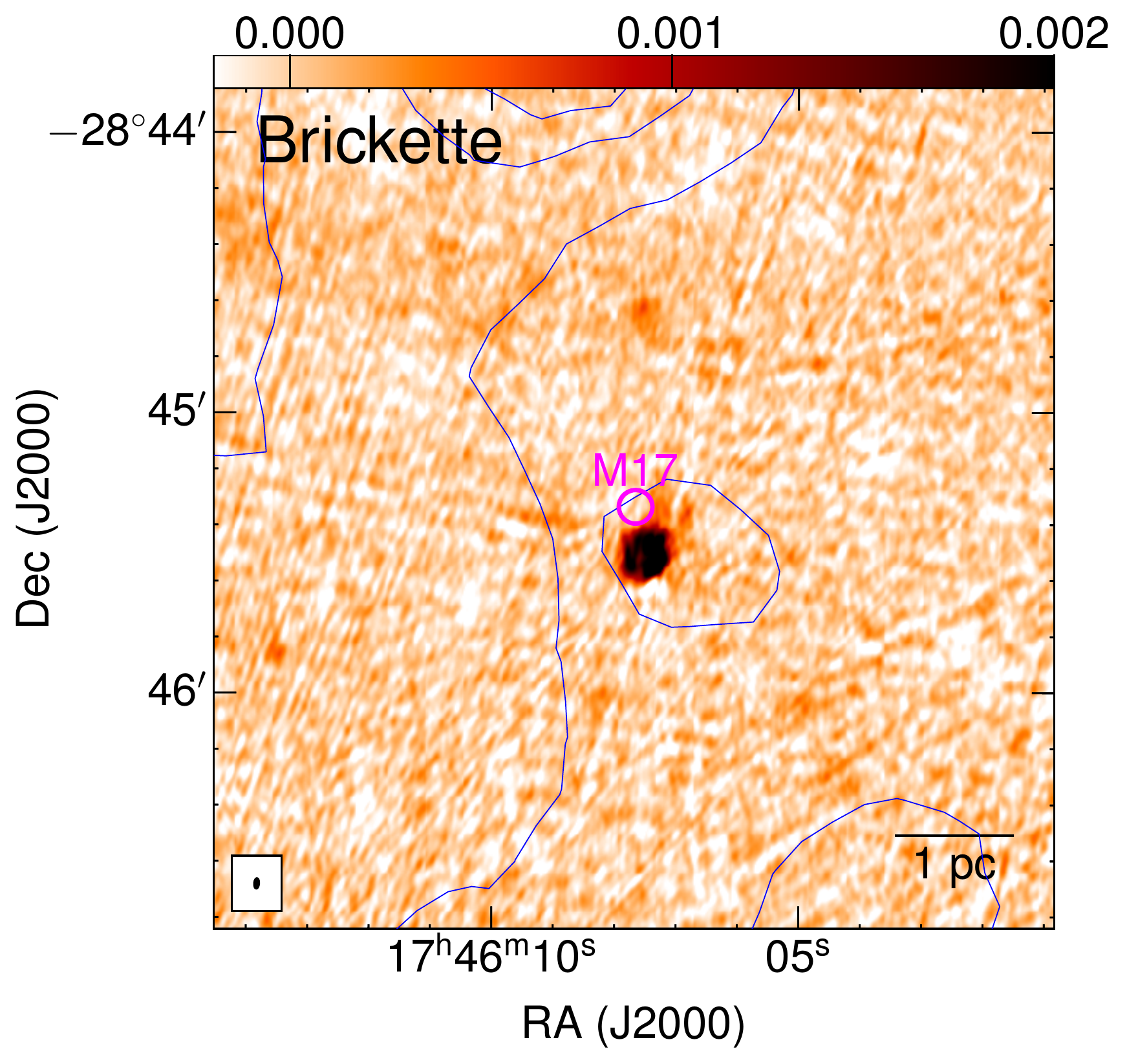} \\
\includegraphics[width=0.38\textwidth]{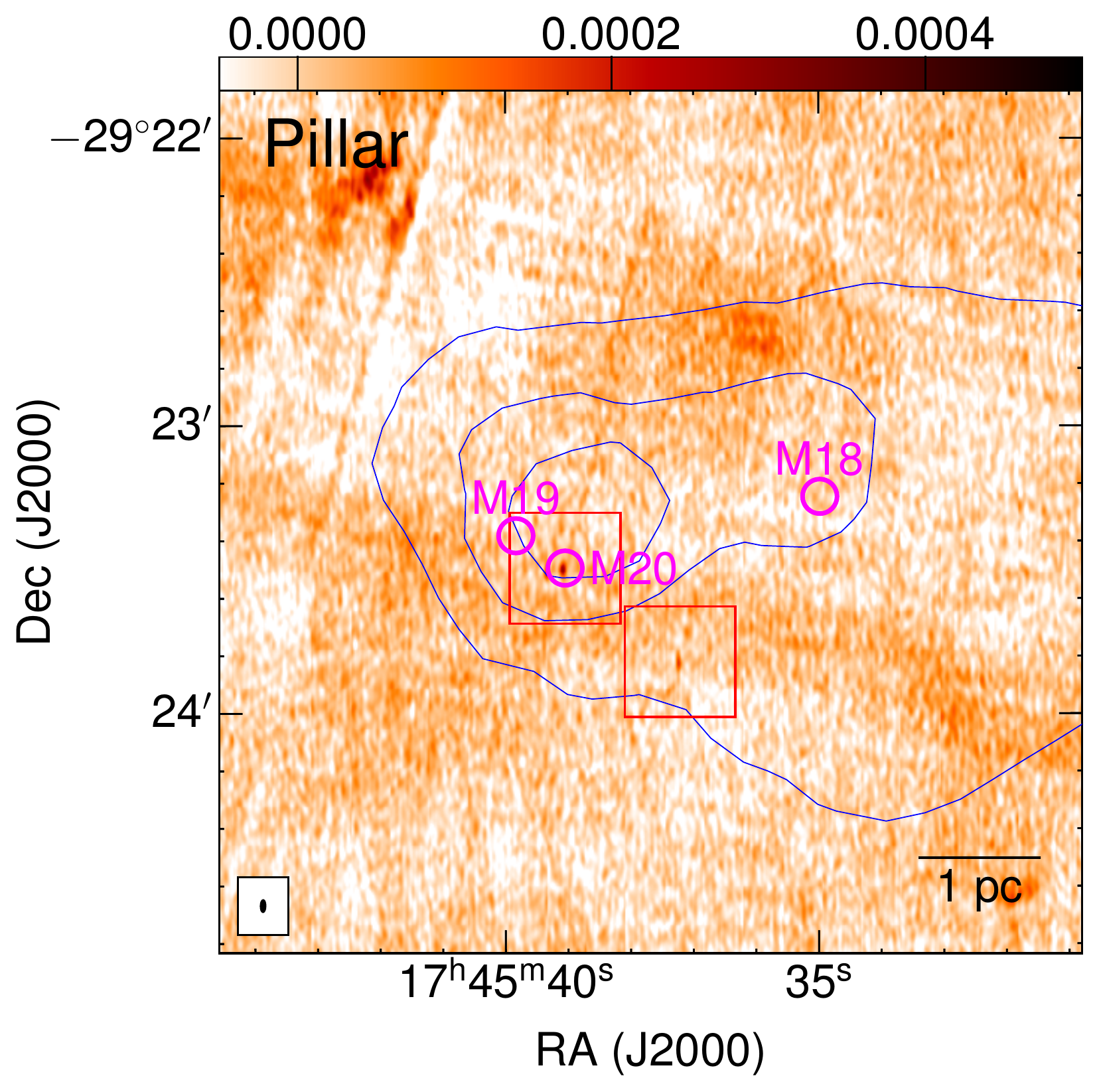} & \includegraphics[width=0.4\textwidth]{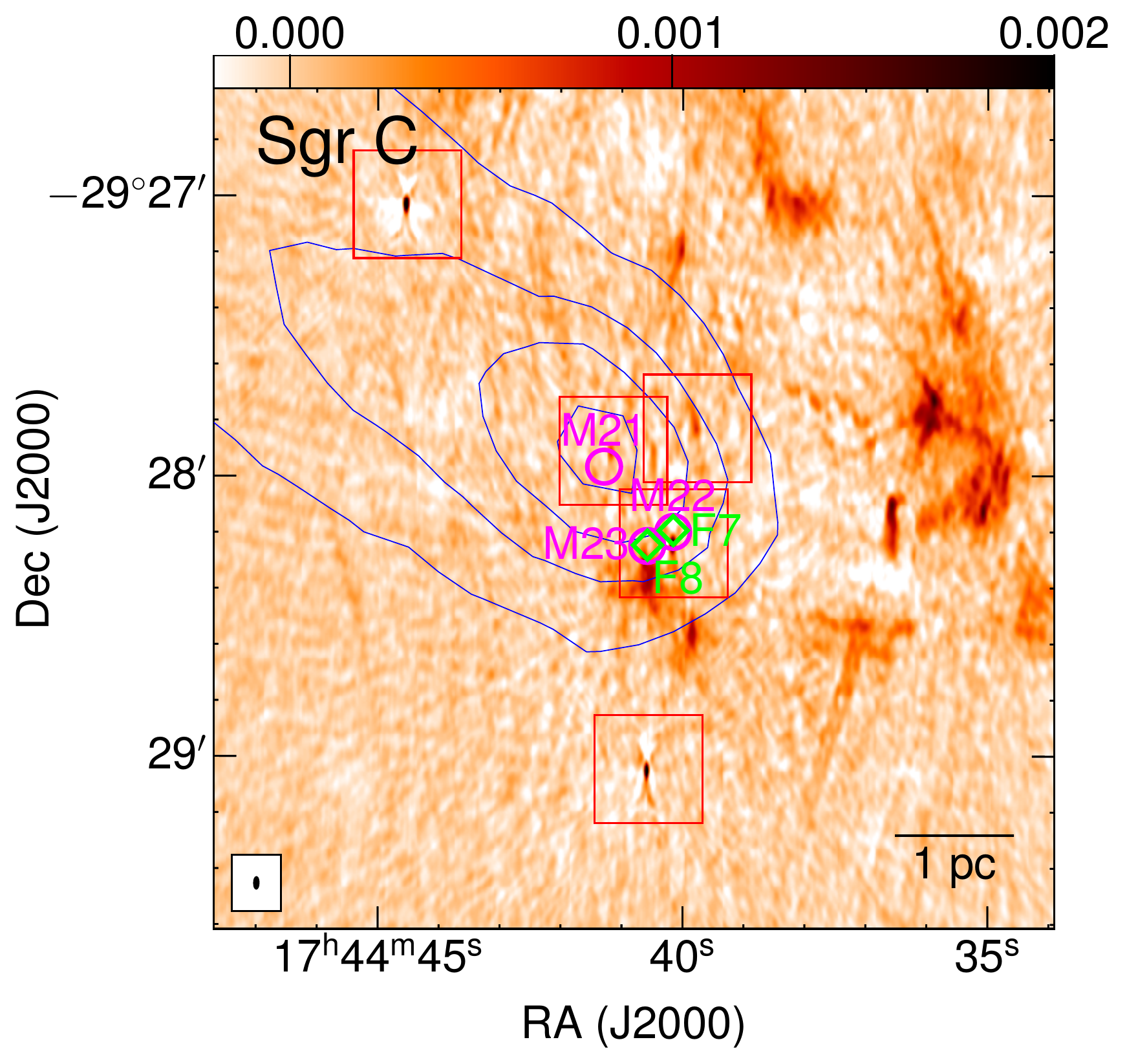}
\end{tabular}
\caption{\meth{} and \fmh{} masers, marked by magenta circles and green diamonds, respectively. Background images show the VLA $C$-band continuum emission. Red boxes mark the compact sources that are identified in continuum (see Figures~\ref{fig:sgrb2}--\ref{fig:sgrc}).}
\label{fig:maser}
\end{figure*}

\begin{figure*}[!t]
\centering
\begin{tabular}{p{0.5\textwidth}p{0.5\textwidth}}
\includegraphics[width=0.5\textwidth]{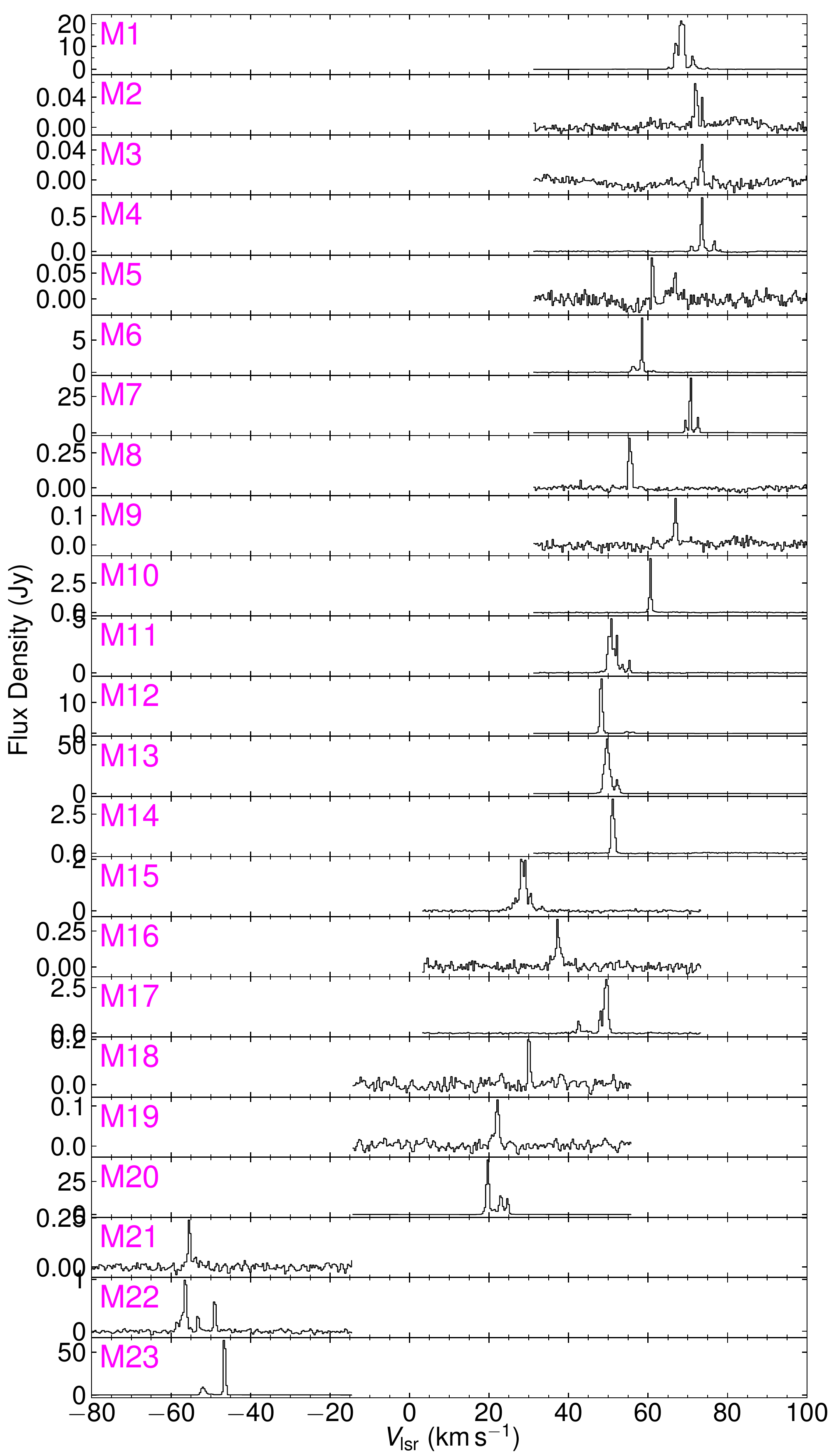} & \includegraphics[width=0.5\textwidth]{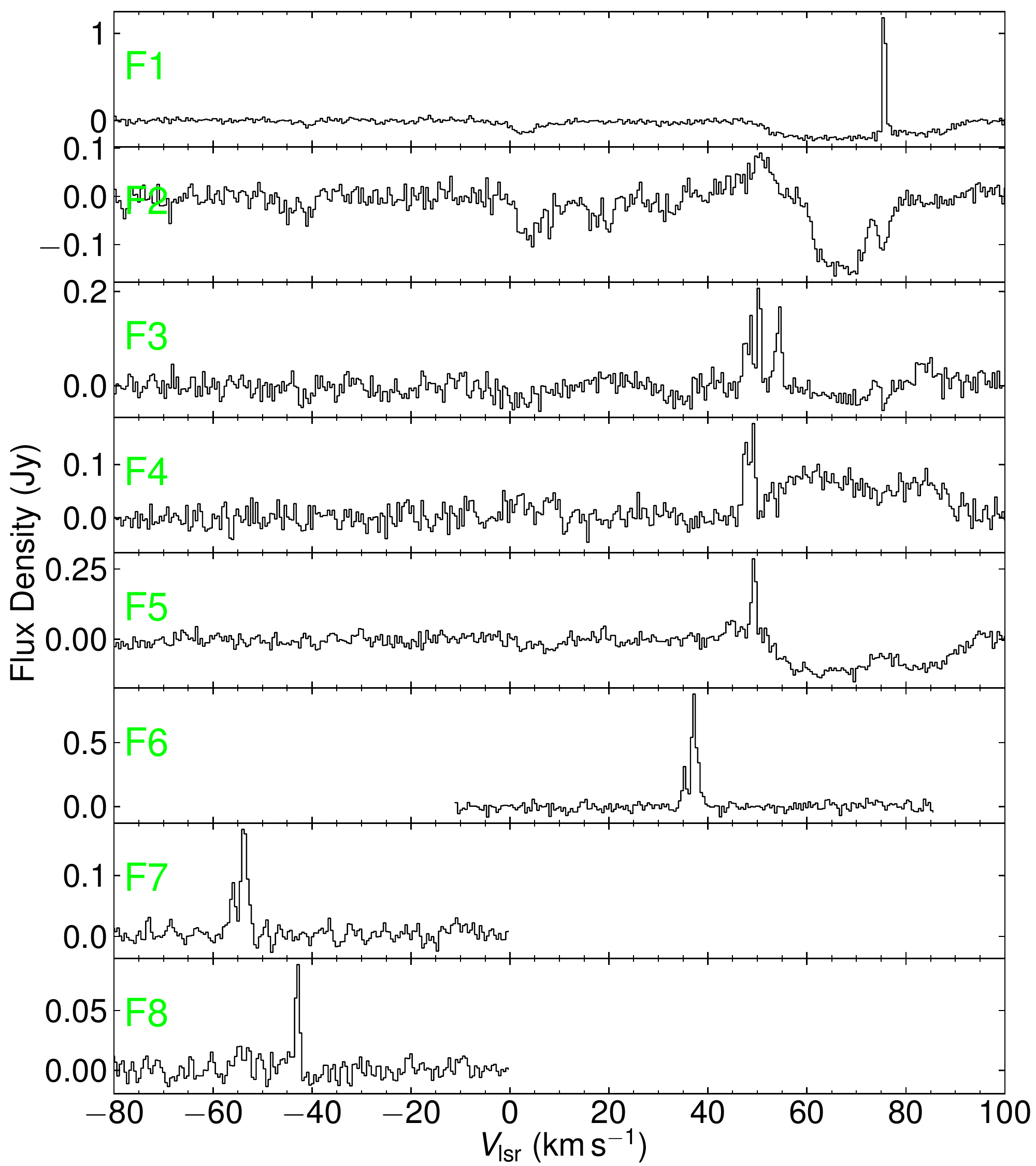}
\end{tabular}
\caption{Spectra of the \meth{} and \fmh{} masers.}
\label{fig:maserspec}
\end{figure*}

We manually identified 6.668~GHz \meth{} masers in the images. After self calibration, the dynamic range was significantly improved, and the images were able to achieve the thermal noise level ($\sim$8~\mjypbm{} in 0.35~\kms{}) except for a few channels in the Sgr~B2 tile. Therefore, identification of masers is straightforward with a visual inspection of maximum intensity maps (the eighth moment maps as defined in CASA). We defined the detection level for maser sources to be above the 5$\sigma$ RMS noise level and be found in at least two channels, where 1$\sigma$ equals 8~\mjypbm{} per 0.35~\kms{} channel. The corresponding brightness temperature criterion is above 10$^3$~K. Therefore, any identified emission should be non-thermal, and as such should be masers, given typical gas temperatures of $\lesssim$300~K in the CMZ \citep{ao2013,ginsburg2016,lu2017,krieger2017}. Multiple velocity components along the same line of sight were classified as a single maser. In the end, we identified 23 masers. Their positions are marked in \autoref{fig:maser} and spectra are shown in \autoref{fig:maserspec}, and properties listed in \autoref{tab:ch3ohmaser}.

All the masers are spatially associated with molecular clouds in the CMZ or in the foreground. 14 masers are detected toward Sgr~B2. Eleven of them have been reported in \citet{houghton1995} and \citet{caswell1996}, and three (M2, M3, and M9) are new detections. Three masers are detected in the Dust Ridge clouds: M15 is detected toward the Dust Ridge cloud e, which has been reported in \citet{caswell2009}; M16 is detected toward Dust Ridge cloud c, which has been reported in \citet{caswell1996}; and M17 is spatially adjacent to an \hii{} region (`Brickette'; previous identified by e.g., \citealt{giveon2005a,immer2012b,rodriguez2013}), which is also known \citep{caswell1996}. Three masers are detected toward Sgr~C, among which two (M22, M 23) have been reported in \citet{caswell1996} and one (M21) is a new detection. Finally, three masers are detected toward a cloud in field 27 (`Pillar') that has been suggested to be in the foreground: the Gaia satellite measured a parallax of 1.3461~miliarcsec toward a bright source spatially coincident with the maser M20 \citep{gaia2018}, which translates to a distance of $\sim$740~pc. Among the three masers detected in this cloud, M20 is known \citep{caswell1996} and two (M18, M19) are new detections. In summary, 17 masers have been reported in the literature and six are new detections. We will discuss the implication for star formation in CMZ molecular clouds in \autoref{subsec:disc_sf}.

Note that toward several positions in Sgr~B2, the 6.668~GHz \meth{} line also shows absorption (e.g., M3 and M5, \autoref{fig:maserspec}), but it is of much smaller magnitude compared with the maser emission, and therefore does not affect our identification of \meth{} masers.

\subsection{\fmh{} Masers}\label{subsec:results_h2co}
\fmh{} emission above the 5$\sigma$ level is only detected toward Sgr~B2, Dust Ridge cloud c, and Sgr~C. Therefore, we manually identify \fmh{} masers in the images of just these three regions. In Dust ridge cloud c and Sgr C, \fmh{} is only detected in emission, thus the identification of masers is straightforward. However in Sgr B2 there is significant absorption which could substantially reduce the peak intensity of the maser emission.

With this consideration in mind, we checked the \fmh{} image of Sgr~B2 channel by channel, and identified point sources above the 5$\sigma$ level as compared to the surrounding continuum level (which could be negative, if affected by absorption) and detected in at least two channels. We found five \fmh{} maser candidates, as shown in \autoref{fig:maser} and \autoref{fig:maserspec}, and three of them are indeed spatially coincident with strong \fmh{} absorption (F1, F2, and F5 in \autoref{fig:maserspec}). All five \fmh{} masers have been reported in \citet{mehringer1994}. Note that F2 shows a complex spectrum with strong emission and absorption features, among which we identified the channels at $\sim$50~\kms{} as maser emission, in line with the finding of \citet{mehringer1994}.

As for Dust Ridge cloud c and Sgr~C, we identified three \fmh{} masers, as shown in \autoref{fig:maser} and \autoref{fig:maserspec}. The one in Dust Ridge cloud c has been reported in \citet{ginsburg2015}. The two \fmh{} masers in Sgr~C are new detections. Properties of the \fmh{} masers are listed in \autoref{tab:h2comaser}.

%%%%%%%%%%%%%%%%%%%%%%%%%
\section{DISCUSSION}\label{sec:discussion}

\subsection{Identification of UC \hii{} Regions}\label{subsec:disc_uchii}
Five of the identified compact sources are spatially associated with \clt{} \meth{} masers (C23, C82, C83, C102, and C103; see Sections~\ref{subsec:results_ch3oh}, \ref{subsec:results_h2co}, \autoref{fig:maser}), and are deemed to be UC \hii{} regions. For the other compact sources, we attempt to investigate their nature by using $C$-band spectral indices and correlations with the hydrogen recombination line emission, infrared emission, high column densities, star clusters and massive field stars, evolved stars, and X-ray sources.

\subsubsection{$C$-band Spectral Indices}\label{subsubsec:disc_spx}
Thermal free-free emission from UC \hii{} regions usually present $\alpha=-0.1$ at $\lesssim$10~GHz, while it could become optically thick in HC \hii{} regions and present a higher spectral $\alpha$ of up to 2 \citep{sanchezmonge2013a}. As shown in \autoref{tab:uchii}, after taking into account the fitting uncertainties and the systematic uncertainty (0.05), 25 out of the 104 sources present spectral indices between $-$0.1 and 2.

However, there are several caveats in interpreting the derived spectral indices, which prevent us from confirming the nature of the compact sources.

The in-band spectral indices derived from our $C$-band data may be biased, and those derived with a wider frequency range may be different. For example, several known UC \hii{} regions present spectral indices of $<$$-$0.1 in our data even after taking the uncertainties into account: C102 and C103 are embedded in the Sgr~C cloud and have been identified as UC \hii{} regions using $K$-band continuum at 23~GHz \citep[labelled as H1 and H3 in Sgr~C;][]{lu2019a}.  Both of them are associated with \clt{} \meth{} masers and C103 is also associated with a \fmh{} maser (Sections~\ref{subsec:results_ch3oh}, \ref{subsec:results_h2co}). Their $C$-band spectral indices are $-$0.35 and $-$0.53, respectively. However, if we estimate multi-band spectral indices using the $C$-band and $K$-band data (assuming 5\% and 10\% flux uncertainties, respectively), the results are 0.34$\pm$0.08 and $-$0.15$\pm$0.08, respectively, and are consistent with thermal free-free emission from UC/HC \hii{} regions. The other case is the source C40 \citep[also known as N3;][]{yusefzadeh1987a}, which has a spectral index of 0.77$\pm$0.01 in our measurement. However, it has been observed in multiple bands with the VLA \citep{ludovici2016}, and a broken power-law spectral profile is found---at low frequencies (2--6~GHz), its spectral index is 0.56$\pm$0.13, which is in agreement with our result within the uncertainties, but at high frequencies (10--36~GHz), the spectral index is $-$0.86$\pm$0.11. As a result, it was ruled out to be an \hii{} region by \citet{ludovici2016}, and could be a background AGN as suggested by Butterfield et al.\ (submitted).

Therefore, the $C$-band spectral indices alone are insufficient to confirm the nature of the sources. In the next section, we try to correlate the compact sources with Pa-$\alpha$ emission, infrared surveys, high column densities, and X-ray observations, to more robustly search for UC \hii{} region candidates.

\subsubsection{Correlation with Other Surveys that Argue for UC \hii{} Regions}\label{subsubsec:disc_others}
First, we compare the identified compact sources with the Hubble Space Telescope (HST) Paschen-$\alpha$ survey toward the inner part of the CMZ \citep[see the black contour in Figures~\ref{fig:dustridge}--\ref{fig:sgrc};][]{wang2010,dong2011}. The detection of Pa-$\alpha$ emission toward a compact source suggests thermal emission, which could be from UC \hii{} regions but could also arise from evolved stars and planetary nebulae (see discussion in \autoref{subsubsec:disc_ruleout}).

\citet{dong2011} compiled Pa-$\alpha$ emitting source catalogs. Sources in the catalogs are point-like with a resolution of 0\farcs{2} from the HST observations, and are mostly evolved high-mass stars as suggested by the authors. The Pa-$\alpha$ emission associated with our compact sources tends to be diffuse and is not cataloged by \citet{dong2011}. Therefore, we compare directly with the Pa-$\alpha$ image instead of the catalogs. Among the 104 sources identified in \autoref{subsec:results_compact}, 28 are within the observed area of HST. Sixteen sources have Pa-$\alpha$ emission counterparts. Nine of them have spectral indices $<$$-$0.1 after taking systematic and fitting uncertainties into account. This again demonstrates the limitation of using spectral indices to infer the nature of the continuum sources.

Second, we compare with \hii{} region catalogs, which are mostly based on infrared emission and in some cases with supplemental radio continuum data. \citet{giveon2005a} have compiled a catalog based on MAGPIS 5~GHz continuum emission \citep[4\arcsec{} resolution;][]{becker1994} and Midcourse Space Experiment (MSX) infrared emission. However, after cross-matching with our $C$-band images, we find that all of their identified \hii{} regions are diffuse structures (e.g., Sgr~B2(M)), and none matches with the compact sources we identify. The other \hii{} region catalog compiled by \citet{anderson2014} makes use of the Wide-Field infrared Survey Explorer (WISE) data ($\sim$6\arcsec{} resolution at the shorter wavelength bands) and is suggested to be the most complete one to date. We search for \hii{} regions in the catalog within a radius of 6\arcsec{} around our compact sources, and find three matches, which are marked in \autoref{tab:uchii}.

Third, we compare with the YSO catalogs of \citet{yusefzadeh2009} and \citet{an2011}, both based on the \textit{Spitzer} infrared data. The \citet{yusefzadeh2009} catalog used 24~\micron{} data ($\sim$6\arcsec{} resolution) to identify 559 YSO candidates, but may be contaminated by more evolved object such as main sequence stars \citep{koepferl2015}, and several candidates were indeed ruled out later by \citet{an2011} through infrared spectra (e.g., the one associated with C75, SSTGC374813). It also includes 33 Extended Green Objects (EGOs), a class of objects that are supposed to be associated with massive YSOs \citep{cyganowski2008}. We cross-match between the YSO candidates and our compact source catalog with a search radius of 6\arcsec{}, and find that eight sources (C10, C23, C56, C63, C72, C75,  C89, and C97) are coincident with YSO candidates. The EGOs are usually spatially extended, so we use a larger search radius of 9\arcsec{} \citep[characteristic radius of EGOs;][]{cyganowski2008} and find four matches (C23, C82/C83, and C103), which have been classified as UC \hii{} regions given correlations with \clt{} \meth{} masers. The \citet{an2011} catalog is smaller, with 16 YSOs and 19 possible YSOs that are selected from the \textit{Spitzer} Infrared Array Camera images ($\sim$2\arcsec{} resolution) and then spectroscopically classified. We find only one match with a search radius of 2\arcsec{}: C37 matches with a possible YSO. All these associations are marked in \autoref{tab:uchii}.

Fourth, we compare with the \textit{Herschel} column densities (\citealt{battersby2011}, C.\ Battersby et al.\ in prep.). UC \hii{} regions are deeply embedded in molecular gas, and therefore should be associated with high column densities, although sources of other nature can be projected onto this area by chance. We apply a column density threshold of 5$\times$10$^{22}$~\sqc{} (the lowest contour level in Figures~\ref{fig:sgrb2}--\ref{fig:sgrc}), and search for compact sources above it. This column density threshold is chosen to be the upper limit of the foreground column density toward Sgr~B2 (\citealt{ginsburg2018}; C.\ Battersby et al.\ in prep.), so any emission above it is very likely associated with true gas components in the CMZ while that below it could be in the background. 30 out of the 104 sources are found above the column density threshold and marked in \autoref{tab:uchii}.

\subsubsection{Correlation with Other Surveys that Argue against UC \hii{} Regions}\label{subsubsec:disc_ruleout}
Apart from star formation in molecular clouds, various alternative processes may contribute to the observed compact continuum emission. Here we compare our identified compact sources with studies that target other objects, including field stars, star clusters, and pulsars, to exclude potential contamination from those objects in our detections.

First, the two young massive star clusters in the CMZ, the Arches and the Quintuplet, are known to host high-mass stars with strong stellar winds and free-free emission that can be detected in radio continuum. We compare with the VLA multi-frequency observations of \citet{lang2005}, and find that C41 is likely associated with a stellar member \citep[source AR1 in the Arches cluster;][]{lang2005}.

Second, field stars may ionize surrounding gas and create \hii{} regions. One example is the Sgr~A-H \hii{} regions, scattered in projection between Sgr~A* and the Arches cluster \citep{yusefzadeh1987b,lang2001}, most of which are likely associated with field stars \citep{dong2017,hankins2017}. We compare with the VLA multi-frequency observations of \citet{lang2001} and Stratospheric Observatory for Infrared Astronomy (SOFIA) mid-infrared observations of \citet{hankins2019} towards this area, and find that C44, C45, and C46 are spatially coincident with three known \hii{} regions (H13, H12, and H11) within a radius of 1\arcsec{}, which may be powered by massive field stars. We also compare with Galactic Center-wide studies of massive field stars \citep{mauerhan2010a,mauerhan2010b,dong2015} and find one match, C51, with a star in \citet{mauerhan2010a} within a radius of 1\arcsec{}.

Third, pulsars, X-ray binaries, stellar winds from massive stars, and background active galactic nuclei (AGNs) can contribute to radio continuum emission. These targets are usually seen as point sources in X-ray emission, while UC \hii{} regions may present diffuse X-ray emission \citep[$\gtrsim$0.2~pc, 5\arcsec{} at the distance of the CMZ; e.g.,][]{tsujimoto2006} therefore can be distinguished. We cross match with \textit{Chandra} X-ray point source catalogs of \citet{muno2006,muno2009} and \citet{zhu2018} with a search radius of 1\arcsec{}, and find the following matches: C1, C5, C25, C41, C48, C51, C55, C60, C63, and C86. Among them, C41 and C51 likely originate from stellar winds of massive stars as discussed above. The other eight sources are unlikely to be UC \hii{} regions either given association with X-ray point sources.

Lastly, evolved stars (e.g., planetary nebulae, PNe; Mira variables) have thermal continuum emission that can be detected in radio frequencies. We cross match with the PNe database of \citet{parker2016}, and find that C47 \citep[also known as N1;][]{yusefzadeh1987a} is consistent with a known PN. In addition, C43 is spatially coincident with a Mira variable from the survey of \citet{glass2001}.

Some sources listed above are already unlikely to be UC \hii{} regions given negative spectral indices or the lack of correlations in \autoref{subsubsec:disc_others}, but we do find a few sources that meet two or more criteria in \autoref{subsubsec:disc_others} yet still are of nature other than UC \hii{} regions (e.g., C41, C43, C47, C51, C86). This is expected because massive stars or evolved stars can present infrared and Pa-$\alpha$ emission and positive spectral indices similar to UC \hii{} regions, and may locate adjacent to massive clouds where they are born (especially for the short-lived massive stars), which explain the correlations found in \autoref{subsubsec:disc_others}. We list all counterparts discussed in this section in \autoref{tab:uchii}.

\subsubsection{UC \hii{} Region Candidates and Nature of the Other Sources}\label{subsubsec:disc_finaluchii}
The above results demonstrate that a single criterion (spectral index, Pa-$\alpha$ emission, WISE \hii{} regions, \textit{Spitzer} YSOs, or column densities) is insufficient to determine the nature of a compact source. Therefore, we combine evidence from different observations: if at least two of the criteria discussed in \autoref{subsubsec:disc_others} are met, then the compact source is taken as an UC \hii{} region candidate. In addition, if the source is found to have any counterpart in \autoref{subsubsec:disc_ruleout}, it is immediately excluded to be an UC \hii{} region. As shown in \autoref{tab:uchii}, we have 12 candidates in addition to the five confirmed cases.

Distances of the UC \hii{} candidates are still unknown and they could be in the foreground or background instead of lying within the CMZ. Two cases are C82 and C83, which appear to be associated with the \meth{} maser M20 that is in the foreground (see \autoref{subsec:results_ch3oh}). Observations of recombination lines toward these candidates can both confirm the nature of the continuum emission and yield velocity information that can be used to infer the correlation with the CMZ along the line of sight.

Of the 87 sources that are not identified as UC \hii{} or candidates, 15 have counterparts such as massive cluster or field stars, X-ray point sources (pulsars, X-ray binaries, or AGNs), or evolved stars. The remaining 72 sources are also likely massive field stars, evolved stars, pulsars, or extragalactic sources instead of embedded UC \hii{} regions, given their negative spectral indices and lack of correlation with high column densities. \change{At least one of these sources, C96, presents a double-lobe morphology, which may suggest radio lobes associated with AGNs. A recent VLA 5.5~GHz survey of the GOODS-North field detected 94 sources (including both star forming galaxies and AGNs) in an area of 150~arcmin$^2$ \citep{guidetti2017}, and 3 of them are above the flux threshold of 0.56~mJy (corresponding to five times the median rms of our observations). If this detection rate is representative for background sources of extragalactic origins, in the area of 1100~arcmin$^2$ covered by our observations, 22 extragalactic sources are expected. Therefore, we expect a substantial fraction of the 72 sources to be background galaxies or AGNs.}
Future observations of recombination lines as well as radio continuum at different wavelengths will help clarify their nature.

\subsection{Implications for Star Formation in the CMZ}\label{subsec:disc_sf}

\subsubsection{Class\,\textsc{ii} \meth{} Masers and High-mass Star Formation}\label{subsubsec:disc_methmasers}
The 6.668~GHz \meth{} maser is one of the radiatively excited \clt{} \meth{} masers, which are suggested to uniquely trace high-mass star formation \citep{menten1991a,ellingsen2006,xu2008,breen2013}. Therefore, it is not surprising that most of the 23 detected \meth{} masers are associated with known high-mass star formation regions. Among them, 12 out of the 14 \meth{} masers in Sgr~B2 are associated with UC/HC \hii{} regions \citep{gaume1995,depree2015} or massive YSOs \citep{ginsburg2018}, while M2 and M11 do not have \hii{} region or YSO counterparts within 1\arcsec{} (see next paragraph). The three masers in the Dust Ridge clouds (M15, M16, and M17) are associated with (UC) \hii{} regions or massive YSOs \citep{immer2012b,walker2018,lu2019a}. The three masers in Sgr~C (M12, M22, and M23) are coincident with three (UC) \hii{} regions, respectively \citep{kendrew2013,lu2019a}. M20 in the Pillar cloud is also coincident with a massive YSO traced by an EGO \citep{yusefzadeh2009,chambers2011,chambers2014}, although it is likely in the foreground, not in the CMZ.

Four masers are not clearly associated with any known high-mass star formation activity. One of them is the previously identified maser M11 in Sgr~B2, which is 5\arcsec{} offset from the \hii{} region Sgr~B2 H \citep[or Sgr~B2 South;][]{gaume1995} and is spatially associated with the \fmh{} maser F3. The other three are newly detected---M2 in Sgr~B2, and M18 and M19 in the Pillar cloud which are likely in the foreground together with the YSO associated with M20 given their similar \vlsr{}. Future high resolution ($\lesssim$1\arcsec{}) observations in the radio and submillimeter bands will help to search for gas components associated with these masers.

The 6.668~GHz \meth{} masers are rarer in the CMZ relative to the 22.235~GHz \water{} masers \citep{walsh2011,lu2019a}, and are concentrated in five high-mass star forming regions (excluding the foreground Pillar cloud). For example, in \citet{lu2019a} we detected numerous \water{} masers in a high-mass star forming region, the 20~\kms{} cloud, but so far no \clt{} \meth{} maser has been detected towards this region.

The \meth{} molecule itself is abundant in the CMZ \citep{jones2013}, therefore the relative dearth of the 6.668~GHz \meth{} masers does not stem from chemistry of interstellar gas. It is more likely due to the excitation condition of this maser, which is related to evolutionary phases or protostellar masses of the high-mass star formation activities.

Among the clouds where \clt{} \meth{} maser are detected, Sgr~C, Dust Ridge clouds c/e, and the Brickette cloud have infrared sources embedded in molecular gas and spatially associated with the masers \citep{immer2012b,kendrew2013,walker2018,lu2019a}. For Sgr~B2, the infrared emission is usually saturated in observations, but we find massive YSO or UC \hii{} region counterparts for most of the masers \citep{gaume1995,depree2015,ginsburg2018}. Therefore, these clouds may provide strong radiation from protostars to excite the \meth{} masers. The 20~\kms{} cloud, on the other hand, does not contain embedded infrared sources corresponding to the star formation signatures traced by \water{} masers and dense cores \citep{lu2019a}. It may be at an even earlier evolutionary phase, or may not harbor protostars above a certain mass threshold that have strong enough radiation to excite \clt{} \meth{} masers.

\subsubsection{\fmh{} Masers and High-mass Star Formation}\label{subsubsec:disc_fmhmasers}
The 4.830~GHz \fmh{} maser has been detected in eight locations in the Galaxy, all of which are high-mass star forming regions \citep[][and references therein]{ginsburg2015}. Our observations reveal Sgr~C (specifically, the two (UC) \hii{} regions in it) as the ninth region with \fmh{} maser detection, which is also a high-mass star forming region. This may suggest that the \fmh{} maser is exclusively associated with high-mass star formation, same as the \clt{} \meth{} maser.

Except for one \fmh{} maser F4, the \fmh{} masers we detect are always projected within 1\arcsec{} around \clt{} \meth{} masers or associated with the same (UC) \hii{} region. This high frequency of co-existence between the two types of masers suggests the excitation condition of the \fmh{} maser is similar to but is more stringent than that of the \clt{} \meth{} maser. For example, the luminosity threshold of YSOs to excite the \fmh{} maser may be higher, or the time period that allows the excitation of the \fmh{} maser may be shorter. Exactly how the \fmh{} maser is excited is still unclear \citep[e.g.,][]{vanderwalt2014}.

\citet{ginsburg2015} discussed the relative prevalence of \fmh{} masers in the CMZ as compared to the Galactic disk, and our new detections reinforce this statement (three regions in the CMZ vs.\ six in the Galactic disk). As suggested by \citet{ginsburg2015}, this may suggest that the \fmh{} masers trace a very short period in high-mass star formation, and the three occurrences of \fmh{} masers in the CMZ indicate an ongoing burst of star formation, or this may be related to high abundance of gas phase \fmh{} at small spatial scales ($\sim$100 AU) in the CMZ.

\subsubsection{Inefficient High-mass Star Formation in the CMZ}\label{subsubsec:disc_inactivesf}
Overall, we find evidence of early phase high-mass star formation traced by \clt{} \meth{} masers and embedded UC \hii{} regions only toward five isolated regions in the inner CMZ: Sgr~B2, Sgr~C, Dust Ridge cloud c, Dust Ridge cloud e, and the Brickette. All of them have been known to form high-mass stars \citep{immer2012b,kendrew2013,walker2015,walker2018,ginsburg2018,lu2019a}. There are another two high-mass star forming clouds, \ctw{} and \cfi{} \citep{mills2011,lu2015b,lu2019a}, where we do not detect any \clt{} \meth{} masers or UC \hii{} regions. Therefore, we are not able to confirm any new high-mass star forming clouds in the CMZ through our observations, and all the currently known high-mass star forming regions in the inner CMZ are confined in seven clouds. \change{A brief summary of the star formation indicators can be found in \autoref{tab:sfsummary}.}

\citet{krumholz2008} suggested a column density threshold for high-mass star formation of 2$\times$10$^{23}$~\sqc{}. Among the seven high-mass star forming clouds, Dust Ridge cloud c, the Brickette, and \cfi{} have column densities below the threshold in the \textit{Herschel} maps (Figures~\ref{fig:dustridge} \& \ref{fig:sgra}), although smaller regions embedded in them above the threshold have been found \citep[e.g.,][]{walker2018,lu2019a}. There are three clouds lying above the threshold but showing no signatures of high-mass star formation in our observations and previous studies: \gzp{}, and Dust Ridge clouds d and f. \gzp{} has been proven to be genuinely lacking star formation \citep{longmore2013b,kauffmann2013a,rathborne2015,mills2015}, while Dust Ridge clouds d and f do not show signs of active on-going star formation either \citep{walker2018,lu2019a,barnes2019}. The strong solenoidally driven turbulence may increase the density threshold for star formation, and therefore inhibit star formation in the three clouds despite their high column densities \citep{federrath2016,henshaw2019,kruijssen2019,dale2019}.

If we simply consider the total molecular gas mass in the inner CMZ of $\sim$10$^7$~\msol{} with a mean density of $\sim$10$^4$~\cc{} \citep{longmore2013a}, the expected SFR based on the dense gas star formation relation extrapolated from nearby clouds \citep{lada2010} is 0.46~\msolpyr{}. Further assuming a typical time scale of 0.3~Myr for the star formation activities traced by UC \hii{} regions and \clt{} \meth{} masers \citep{davies2011}, and a canonical multiple-power-law initial mass function between 0.01~\msol{} and 150~\msol{} \citep{kroupa2001}, we expect to find between 940 and 1069 high-mass protostars above 10--11~\msol{} in this region \citep[see Appendix~D of][]{lu2019a}. Here the threshold of 10--11~\msol{} corresponds to the detection limit of the continuum emission (see \autoref{subsec:results_compact}). However, even after taking into account maser variability or multiplicity of protostars that may result in an underestimate of a factor 2 in observations of star formation indicators, this still outnumbers the observed UC \hii{} regions and \clt{} \meth{} masers combined in the inner CMZ by an order of magnitude (58 in total, see \autoref{tab:sfsummary}). Alternatively, we compare the SFR based on the observed star formation indicators with that expected by the dense gas star formation relation. We assume each UC \hii{} region or \clt{} \meth{} maser corresponds to a high-mass protostar above 10~\msol{}, and follow the methods in Appendix~D of \citet{lu2019a} to estimate the SFR. The results are listed in \autoref{tab:sfsummary}, and the total SFR in the surveyed area, 0.025~\msolpyr{}, is an order of magnitude smaller than 0.46~\msolpyr{} that is expected by the dense gas star formation relation. Our observations therefore strengthen the conclusion that star formation in the CMZ is suppressed by about a factor of 10 than expected by the dense gas star formation relation, which has been drawn from observations of more evolved phases of star formation \citep[in the last several Myr;][]{longmore2013a,barnes2017}. Furthermore, since our observations trace very early phases of star formation deeply embedded in molecular clouds, the results imply that the incipient star formation (in the last $\sim$0.3~Myr) in the CMZ remains to be inefficient, and that any impending, next burst of star formation has not yet begun \citep{KK2015}.

Finally, in spite of the overall inefficient star formation in the CMZ, we confirm that Sgr~C is actively forming stars. Our new detections of \fmh{} masers toward Sgr~C make it one of the most maser-rich regions in the Galaxy, similar to Sgr~B2 and Dust Ridge cloud c. So far, \water{} \citep{caswell1983,walsh2011}, OH \citep{caswell1998,cotton2016}, \clt{} \meth{}, and \fmh{} masers have been detected toward the two (UC) \hii{} regions in Sgr~C. Outside of the two (UC) \hii{} regions, a total of 14 \water{} masers are detected throughout the Sgr~C cloud \citep{lu2019a}, one of which is spatially coincident with the the \clt{} \meth{} maser M21 (the \water{} maser W8, see Figure~4 of \citealt{lu2019a}). Therefore, at least three positions in Sgr~C are forming high-mass stars, creating a variety of masers and UC \hii{} regions, while more than ten other positions are likely forming low to intermediate-mass stars. In fact, Sgr~C is one of the few CMZ clouds that show SFRs consistent with the dense gas star formation relation, along with Sgr~B2 and Dust Ridge cloud c \citep{kauffmann2017a,lu2019a}. The active star formation in Sgr~C may be related to its high fragmentation level and large fraction of gas mass confined in gravitationally bound cores, as opposed to the lack of fragmentation in most other CMZ clouds (e.g., \citealt{kauffmann2017b}, C.~Battersby et al.\ in prep., H.~Hatchfield et al.\ in prep.), though the origin of this unique gas structure is unclear. It may be a combined effect of self-gravity, impact of a nearby 10-pc scale \hii{} region \citep[e.g., gas collapse triggered by expanding ionization fonts of the \hii{} region;][]{liszt1995,lang2010}, and global gas dynamics in the CMZ \citep[e.g., gas compression induced by the tidal field of the CMZ;][]{kruijssen2015,kruijssen2019,jeffreson2018,dale2019}.

%%%%%%%%%%%%%%%%%%%%%%%%%
\section{CONCLUSIONS}\label{sec:conclusions}
We report new VLA observations of $C$-band continuum emission, 6.668~GHz \meth{} maser, and 4.830~GHz \fmh{} maser at $\sim$1\arcsec{} resolution toward the inner part of the CMZ. We use these data to search for high-mass star formation at early evolutionary phases in the CMZ. The continuum observation is complete to free-free emission from stars above 10--11~\msol{} throughout the inner 200~pc of the CMZ except for small regions around bright continuum sources in Sgr~B2 and SgrA. Using the continuum emission, we confirm 5 UC \hii{} regions, and find 12 UC \hii{} region candidates whose nature needs to be verified in the future. We detect 23 \meth{} masers and eight \fmh{} masers, among which six and two are new detections, respectively.

Despite the new candidates of UC \hii{} regions and the new detections of masers, we do not find more signatures of on-going high-mass star formation than previously known in the CMZ. Our observations suggest that current high-mass star formation in the CMZ is concentrated in a few isolated regions with high column densities ($\gtrsim$10$^{23}$~\sqc{}, including \ctw{}, \cfi{}, Dust Ridge clouds c and e, Brickette, Sgr B2, and Sgr C). Combined with previous studies that focus on more evolved phases (in the last several Myr) of star formation in the CMZ and find a star formation efficiency at least 10 times lower than expected by the dense gas star formation relation, our results indicate that star formation at early evolutionary phases (in the last $\sim$0.3~Myr) in the CMZ remains to be inefficient, and that if there will be any impending, next burst of star formation, it has not yet begun.

\acknowledgments
\change{We thank the referee, Daniel Wang, for helpful comments.} We thank NRAO staff for their support with the VLA observations and data reduction. This work was supported by JSPS KAKENHI grant No.\ JP18K13589. CB was supported by the National Science Foundation under Grant No.\ 1816715. JMDK gratefully acknowledges funding from the German Research Foundation (DFG) in the form of an Emmy Noether Research Group (grant number \mbox{KR4801/1-1}) and from the European Research Council (ERC) under the European Union's Horizon 2020 research and innovation programme via the ERC Starting Grant MUSTANG (grant agreement number 714907). This research made use of Astropy, a community-developed core Python package for Astronomy \citep{astropy2013}, and APLpy, an open-source plotting package for Python \citep{aplpy2012}. Data analysis was in part carried out on the open use data analysis computer system at the Astronomy Data Center (ADC) of the National Astronomical Observatory of Japan. This research has made use of NASA’s Astrophysics Data System. This research has made use of the SIMBAD database and the VizieR catalogue access tool, operated at CDS, Strasbourg, France.

\facilities{VLA}

\software{CASA \citep{mcmullin2007}, SExtractor \citep{bertin1996}, BLOBCAT \citep{hales2012}, APLpy \citep{aplpy2012}, Astropy \citep{astropy2013}}

\bibliographystyle{aasjournal}
\bibliography{HMSFinCMZ_forarxiv.bbl}

\begin{thebibliography}{}
\expandafter\ifx\csname natexlab\endcsname\relax\def\natexlab#1{#1}\fi
\providecommand{\url}[1]{\href{#1}{#1}}
\providecommand{\dodoi}[1]{doi:~\href{http://doi.org/#1}{\nolinkurl{#1}}}
\providecommand{\doeprint}[1]{\href{http://ascl.net/#1}{\nolinkurl{http://ascl.net/#1}}}
\providecommand{\doarXiv}[1]{\href{https://arxiv.org/abs/#1}{\nolinkurl{https://arxiv.org/abs/#1}}}

\bibitem[{{An} {et~al.}(2011){An}, {Ram{\'{\i}}rez}, {Sellgren}, {Arendt},
  {Adwin Boogert}, {Robitaille}, {Schultheis}, {Cotera}, {Smith}, \&
  {Stolovy}}]{an2011}
{An}, D., {Ram{\'{\i}}rez}, S.~V., {Sellgren}, K., {et~al.} 2011, \apj, 736,
  133, \dodoi{10.1088/0004-637X/736/2/133}

\bibitem[{{Anderson} {et~al.}(2014){Anderson}, {Bania}, {Balser}, {Cunningham},
  {Wenger}, {Johnstone}, \& {Armentrout}}]{anderson2014}
{Anderson}, L.~D., {Bania}, T.~M., {Balser}, D.~S., {et~al.} 2014, \apjs, 212,
  1, \dodoi{10.1088/0067-0049/212/1/1}

\bibitem[{{Ao} {et~al.}(2013){Ao}, {Henkel}, {Menten}, {Requena-Torres},
  {Stanke}, {Mauersberger}, {Aalto}, {M{\"u}hle}, \& {Mangum}}]{ao2013}
{Ao}, Y., {Henkel}, C., {Menten}, K.~M., {et~al.} 2013, \aap, 550, A135,
  \dodoi{10.1051/0004-6361/201220096}

\bibitem[{{Astropy Collaboration} {et~al.}(2013){Astropy Collaboration},
  {Robitaille}, {Tollerud}, {Greenfield}, {Droettboom}, {Bray}, {Aldcroft},
  {Davis}, {Ginsburg}, {Price-Whelan}, {Kerzendorf}, {Conley}, {Crighton},
  {Barbary}, {Muna}, {Ferguson}, {Grollier}, {Parikh}, {Nair}, {Unther},
  {Deil}, {Woillez}, {Conseil}, {Kramer}, {Turner}, {Singer}, {Fox}, {Weaver},
  {Zabalza}, {Edwards}, {Azalee Bostroem}, {Burke}, {Casey}, {Crawford},
  {Dencheva}, {Ely}, {Jenness}, {Labrie}, {Lim}, {Pierfederici}, {Pontzen},
  {Ptak}, {Refsdal}, {Servillat}, \& {Streicher}}]{astropy2013}
{Astropy Collaboration}, {Robitaille}, T.~P., {Tollerud}, E.~J., {et~al.} 2013,
  \aap, 558, A33, \dodoi{10.1051/0004-6361/201322068}

\bibitem[{{Bally} {et~al.}(1987){Bally}, {Stark}, {Wilson}, \&
  {Henkel}}]{bally1987}
{Bally}, J., {Stark}, A.~A., {Wilson}, R.~W., \& {Henkel}, C. 1987, \apjs, 65,
  13, \dodoi{10.1086/191217}

\bibitem[{{Barnes} {et~al.}(2017){Barnes}, {Longmore}, {Battersby}, {Bally},
  {Kruijssen}, {Henshaw}, \& {Walker}}]{barnes2017}
{Barnes}, A.~T., {Longmore}, S.~N., {Battersby}, C., {et~al.} 2017, \mnras,
  469, 2263, \dodoi{10.1093/mnras/stx941}

\bibitem[{{Barnes} {et~al.}(2019){Barnes}, {Longmore}, {Avison}, {Contreras},
  {Ginsburg}, {Henshaw}, {Rathborne}, {Walker}, {Alves}, \&
  {Bally}}]{barnes2019}
{Barnes}, A.~T., {Longmore}, S.~N., {Avison}, A., {et~al.} 2019, \mnras, 486,
  283, \dodoi{10.1093/mnras/stz796}

\bibitem[{{Battersby} {et~al.}(2011){Battersby}, {Bally}, {Ginsburg},
  {Bernard}, {Brunt}, {Fuller}, {Martin}, {Molinari}, {Mottram}, {Peretto},
  {Testi}, \& {Thompson}}]{battersby2011}
{Battersby}, C., {Bally}, J., {Ginsburg}, A., {et~al.} 2011, \aap, 535, A128,
  \dodoi{10.1051/0004-6361/201116559}

\bibitem[{{Becker} {et~al.}(1994){Becker}, {White}, {Helfand}, \&
  {Zoonematkermani}}]{becker1994}
{Becker}, R.~H., {White}, R.~L., {Helfand}, D.~J., \& {Zoonematkermani}, S.
  1994, \apjs, 91, 347, \dodoi{10.1086/191941}

\bibitem[{{Bertin} \& {Arnouts}(1996)}]{bertin1996}
{Bertin}, E., \& {Arnouts}, S. 1996, \aaps, 117, 393,
  \dodoi{10.1051/aas:1996164}

\bibitem[{{Bihr} {et~al.}(2016){Bihr}, {Johnston}, {Beuther}, {Anderson},
  {Ott}, {Rugel}, {Bigiel}, {Brunthaler}, {Glover}, {Henning}, {Heyer},
  {Klessen}, {Linz}, {Longmore}, {McClure-Griffiths}, {Menten}, {Plume},
  {Schierhuber}, {Shanahan}, {Stil}, {Urquhart}, \& {Walsh}}]{bihr2016}
{Bihr}, S., {Johnston}, K.~G., {Beuther}, H., {et~al.} 2016, \aap, 588, A97,
  \dodoi{10.1051/0004-6361/201527697}

\bibitem[{{Breen} {et~al.}(2013){Breen}, {Ellingsen}, {Contreras}, {Green},
  {Caswell}, {Stevens}, {Dawson}, \& {Voronkov}}]{breen2013}
{Breen}, S.~L., {Ellingsen}, S.~P., {Contreras}, Y., {et~al.} 2013, \mnras,
  435, 524, \dodoi{10.1093/mnras/stt1315}

\bibitem[{{Butterfield} {et~al.}(2018){Butterfield}, {Lang}, {Morris}, {Mills},
  \& {Ott}}]{butterfield2018}
{Butterfield}, N., {Lang}, C.~C., {Morris}, M., {Mills}, E. A.~C., \& {Ott}, J.
  2018, \apj, 852, 11, \dodoi{10.3847/1538-4357/aa886e}

\bibitem[{{Caswell}(1996)}]{caswell1996}
{Caswell}, J.~L. 1996, \mnras, 283, 606, \dodoi{10.1093/mnras/283.2.606}

\bibitem[{{Caswell}(1998)}]{caswell1998}
---. 1998, \mnras, 297, 215, \dodoi{10.1046/j.1365-8711.1998.01468.x}

\bibitem[{{Caswell}(2009)}]{caswell2009}
---. 2009, PASA, 26, 454, \dodoi{10.1071/AS09013}

\bibitem[{{Caswell} {et~al.}(1983){Caswell}, {Batchelor}, {Forster}, \&
  {Wellington}}]{caswell1983}
{Caswell}, J.~L., {Batchelor}, R.~A., {Forster}, J.~R., \& {Wellington}, K.~J.
  1983, AuJPh, 36, 401, \dodoi{10.1071/PH830401b}

\bibitem[{{Caswell} {et~al.}(2010){Caswell}, {Fuller}, {Green}, {Avison},
  {Breen}, {Brooks}, {Burton}, {Chrysostomou}, {Cox}, {Diamond}, {Ellingsen},
  {Gray}, {Hoare}, {Masheder}, {McClure-Griffiths}, {Pestalozzi}, {Phillips},
  {Quinn}, {Thompson}, {Voronkov}, {Walsh}, {Ward-Thompson}, {Wong-McSweeney},
  {Yates}, \& {Cohen}}]{caswell2010}
{Caswell}, J.~L., {Fuller}, G.~A., {Green}, J.~A., {et~al.} 2010, \mnras, 404,
  1029, \dodoi{10.1111/j.1365-2966.2010.16339.x}

\bibitem[{{Chambers} {et~al.}(2014){Chambers}, {Yusef-Zadeh}, \&
  {Ott}}]{chambers2014}
{Chambers}, E.~T., {Yusef-Zadeh}, F., \& {Ott}, J. 2014, \aap, 563, A68,
  \dodoi{10.1051/0004-6361/201322752}

\bibitem[{{Chambers} {et~al.}(2011){Chambers}, {Yusef-Zadeh}, \&
  {Roberts}}]{chambers2011}
{Chambers}, E.~T., {Yusef-Zadeh}, F., \& {Roberts}, D. 2011, \apj, 733, 42,
  \dodoi{10.1088/0004-637X/733/1/42}

\bibitem[{{Churchwell}(2002)}]{churchwell2002}
{Churchwell}, E. 2002, \araa, 40, 27,
  \dodoi{10.1146/annurev.astro.40.060401.093845}

\bibitem[{{Cotton} \& {Yusef-Zadeh}(2016)}]{cotton2016}
{Cotton}, W.~D., \& {Yusef-Zadeh}, F. 2016, \apjs, 227, 10,
  \dodoi{10.3847/0067-0049/227/1/10}

\bibitem[{{Cyganowski} {et~al.}(2008){Cyganowski}, {Whitney}, {Holden},
  {Braden}, {Brogan}, {Churchwell}, {Indebetouw}, {Watson}, {Babler},
  {Benjamin}, {Gomez}, {Meade}, {Povich}, {Robitaille}, \&
  {Watson}}]{cyganowski2008}
{Cyganowski}, C.~J., {Whitney}, B.~A., {Holden}, E., {et~al.} 2008, \aj, 136,
  2391, \dodoi{10.1088/0004-6256/136/6/2391}

\bibitem[{{Dale} {et~al.}(2019){Dale}, {Kruijssen}, \& {Longmore}}]{dale2019}
{Dale}, J.~E., {Kruijssen}, J.~M.~D., \& {Longmore}, S.~N. 2019, \mnras, 486,
  3307, \dodoi{10.1093/mnras/stz888}

\bibitem[{{Davies} {et~al.}(2011){Davies}, {Hoare}, {Lumsden}, {Hosokawa},
  {Oudmaijer}, {Urquhart}, {Mottram}, \& {Stead}}]{davies2011}
{Davies}, B., {Hoare}, M.~G., {Lumsden}, S.~L., {et~al.} 2011, \mnras, 416,
  972, \dodoi{10.1111/j.1365-2966.2011.19095.x}

\bibitem[{{de Pree} {et~al.}(1995){de Pree}, {Gaume}, {Goss}, \&
  {Claussen}}]{depree1995}
{de Pree}, C.~G., {Gaume}, R.~A., {Goss}, W.~M., \& {Claussen}, M.~J. 1995,
  \apj, 451, 284, \dodoi{10.1086/176218}

\bibitem[{{de Pree} {et~al.}(1996){de Pree}, {Gaume}, {Goss}, \&
  {Claussen}}]{depree1996}
---. 1996, \apj, 464, 788, \dodoi{10.1086/177364}

\bibitem[{{De Pree} {et~al.}(2015){De Pree}, {Peters}, {Mac Low}, {Wilner},
  {Goss}, {Galv{\'a}n-Madrid}, {Keto}, {Klessen}, \& {Monsrud}}]{depree2015}
{De Pree}, C.~G., {Peters}, T., {Mac Low}, M.~M., {et~al.} 2015, \apj, 815,
  123, \dodoi{10.1088/0004-637X/815/2/123}

\bibitem[{{Dong} {et~al.}(2015){Dong}, {Mauerhan}, {Morris}, {Wang}, \&
  {Cotera}}]{dong2015}
{Dong}, H., {Mauerhan}, J., {Morris}, M.~R., {Wang}, Q.~D., \& {Cotera}, A.
  2015, \mnras, 446, 842, \dodoi{10.1093/mnras/stu2116}

\bibitem[{{Dong} {et~al.}(2011){Dong}, {Wang}, {Cotera}, {Stolovy}, {Morris},
  {Mauerhan}, {Mills}, {Schneider}, {Calzetti}, \& {Lang}}]{dong2011}
{Dong}, H., {Wang}, Q.~D., {Cotera}, A., {et~al.} 2011, \mnras, 417, 114,
  \dodoi{10.1111/j.1365-2966.2011.19013.x}

\bibitem[{{Dong} {et~al.}(2017){Dong}, {Lacy}, {Sch{\"o}del}, {Nogueras-Lara},
  {Gallego-Calvente}, {Mauerhan}, {Wang}, {Cotera}, \&
  {Gallego-Cano}}]{dong2017}
{Dong}, H., {Lacy}, J.~H., {Sch{\"o}del}, R., {et~al.} 2017, \mnras, 470, 561,
  \dodoi{10.1093/mnras/stx1266}

\bibitem[{{Ekers} {et~al.}(1983){Ekers}, {van Gorkom}, {Schwarz}, \&
  {Goss}}]{ekers1983}
{Ekers}, R.~D., {van Gorkom}, J.~H., {Schwarz}, U.~J., \& {Goss}, W.~M. 1983,
  \aap, 122, 143

\bibitem[{{Ellingsen}(2006)}]{ellingsen2006}
{Ellingsen}, S.~P. 2006, \apj, 638, 241, \dodoi{10.1086/498673}

\bibitem[{{Federrath} {et~al.}(2016){Federrath}, {Rathborne}, {Longmore},
  {Kruijssen}, {Bally}, {Contreras}, {Crocker}, {Garay}, {Jackson}, {Testi}, \&
  {Walsh}}]{federrath2016}
{Federrath}, C., {Rathborne}, J.~M., {Longmore}, S.~N., {et~al.} 2016, \apj,
  832, 143, \dodoi{10.3847/0004-637X/832/2/143}

\bibitem[{{Forster} \& {Caswell}(2000)}]{forster2000}
{Forster}, J.~R., \& {Caswell}, J.~L. 2000, \apj, 530, 371,
  \dodoi{10.1086/308347}

\bibitem[{{Gaia Collaboration} {et~al.}(2018){Gaia Collaboration}, {Brown},
  {Vallenari}, {Prusti}, {de Bruijne}, {Babusiaux}, {Bailer-Jones}, {Biermann},
  {Evans}, \& {Eyer}}]{gaia2018}
{Gaia Collaboration}, {Brown}, A.~G.~A., {Vallenari}, A., {et~al.} 2018, \aap,
  616, A1, \dodoi{10.1051/0004-6361/201833051}

\bibitem[{{Gaume} {et~al.}(1995){Gaume}, {Claussen}, {de Pree}, {Goss}, \&
  {Mehringer}}]{gaume1995}
{Gaume}, R.~A., {Claussen}, M.~J., {de Pree}, C.~G., {Goss}, W.~M., \&
  {Mehringer}, D.~M. 1995, \apj, 449, 663, \dodoi{10.1086/176087}

\bibitem[{{Ginsburg} {et~al.}(2015){Ginsburg}, {Walsh}, {Henkel}, {Jones},
  {Cunningham}, {Kauffmann}, {Pillai}, {Mills}, {Ott}, {Kruijssen}, {Menten},
  {Battersby}, {Rathborne}, {Contreras}, {Longmore}, {Walker}, {Dawson}, \&
  {Lopez}}]{ginsburg2015}
{Ginsburg}, A., {Walsh}, A., {Henkel}, C., {et~al.} 2015, \aap, 584, L7,
  \dodoi{10.1051/0004-6361/201527452}

\bibitem[{{Ginsburg} {et~al.}(2016){Ginsburg}, {Henkel}, {Ao}, {Riquelme},
  {Kauffmann}, {Pillai}, {Mills}, {Requena-Torres}, {Immer}, {Testi}, {Ott},
  {Bally}, {Battersby}, {Darling}, {Aalto}, {Stanke}, {Kendrew}, {Kruijssen},
  {Longmore}, {Dale}, {Guesten}, \& {Menten}}]{ginsburg2016}
{Ginsburg}, A., {Henkel}, C., {Ao}, Y., {et~al.} 2016, \aap, 586, A50,
  \dodoi{10.1051/0004-6361/201526100}

\bibitem[{{Ginsburg} {et~al.}(2018){Ginsburg}, {Bally}, {Barnes}, {Bastian},
  {Battersby}, {Beuther}, {Brogan}, {Contreras}, {Corby}, {Darling}, {De Pree},
  {Galv{\'a}n-Madrid}, {Garay}, {Henshaw}, {Hunter}, {Kruijssen}, {Longmore},
  {Lu}, {Meng}, {Mills}, {Ott}, {Pineda}, {S{\'a}nchez-Monge}, {Schilke},
  {Schmiedeke}, {Walker}, \& {Wilner}}]{ginsburg2018}
{Ginsburg}, A., {Bally}, J., {Barnes}, A., {et~al.} 2018, \apj, 853, 171,
  \dodoi{10.3847/1538-4357/aaa6d4}

\bibitem[{{Giveon} {et~al.}(2005){Giveon}, {Becker}, {Helfand}, \&
  {White}}]{giveon2005a}
{Giveon}, U., {Becker}, R.~H., {Helfand}, D.~J., \& {White}, R.~L. 2005, \aj,
  129, 348, \dodoi{10.1086/426360}

\bibitem[{{Glass} {et~al.}(2001){Glass}, {Matsumoto}, {Carter}, \&
  {Sekiguchi}}]{glass2001}
{Glass}, I.~S., {Matsumoto}, S., {Carter}, B.~S., \& {Sekiguchi}, K. 2001,
  \mnras, 321, 77, \dodoi{10.1046/j.1365-8711.2001.03971.x}

\bibitem[{{Gravity Collaboration} {et~al.}(2018){Gravity Collaboration},
  {Abuter}, {Amorim}, {Anugu}, {Baub{\"o}ck}, {Benisty}, {Berger}, {Blind},
  {Bonnet}, {Brandner}, {Buron}, {Collin}, {Chapron}, {Cl{\'e}net}, {Coud{\'e}
  Du Foresto}, {de Zeeuw}, {Deen}, {Delplancke-Str{\"o}bele}, {Dembet},
  {Dexter}, {Duvert}, {Eckart}, {Eisenhauer}, {Finger}, {F{\"o}rster
  Schreiber}, {F{\'e}dou}, {Garcia}, {Garcia Lopez}, {Gao}, {Gendron},
  {Genzel}, {Gillessen}, {Gordo}, {Habibi}, {Haubois}, {Haug}, {Hau{\ss}mann},
  {Henning}, {Hippler}, {Horrobin}, {Hubert}, {Hubin}, {Jimenez Rosales},
  {Jochum}, {Jocou}, {Kaufer}, {Kellner}, {Kendrew}, {Kervella}, {Kok},
  {Kulas}, {Lacour}, {Lapeyr{\`e}re}, {Lazareff}, {Le Bouquin}, {L{\'e}na},
  {Lippa}, {Lenzen}, {M{\'e}rand}, {M{\"u}ler}, {Neumann}, {Ott}, {Palanca},
  {Paumard}, {Pasquini}, {Perraut}, {Perrin}, {Pfuhl}, {Plewa}, {Rabien},
  {Ram{\'{\i}}rez}, {Ramos}, {Rau}, {Rodr{\'{\i}}guez-Coira}, {Rohloff},
  {Rousset}, {Sanchez-Bermudez}, {Scheithauer}, {Sch{\"o}ller}, {Schuler},
  {Spyromilio}, {Straub}, {Straubmeier}, {Sturm}, {Tacconi}, {Tristram},
  {Vincent}, {von Fellenberg}, {Wank}, {Waisberg}, {Widmann}, {Wieprecht},
  {Wiest}, {Wiezorrek}, {Woillez}, {Yazici}, {Ziegler}, \&
  {Zins}}]{gravity2018}
{Gravity Collaboration}, {Abuter}, R., {Amorim}, A., {et~al.} 2018, \aap, 615,
  L15, \dodoi{10.1051/0004-6361/201833718}

\bibitem[{{Guidetti} {et~al.}(2017){Guidetti}, {Bondi}, {Prandoni}, {Muxlow},
  {Beswick}, {Wrigley}, {Smail}, {McHardy}, {Thomson}, {Radcliffe}, \&
  {Argo}}]{guidetti2017}
{Guidetti}, D., {Bondi}, M., {Prandoni}, I., {et~al.} 2017, \mnras, 471, 210,
  \dodoi{10.1093/mnras/stx1162}

\bibitem[{{Hales} {et~al.}(2012){Hales}, {Murphy}, {Curran}, {Middelberg},
  {Gaensler}, \& {Norris}}]{hales2012}
{Hales}, C.~A., {Murphy}, T., {Curran}, J.~R., {et~al.} 2012, \mnras, 425, 979,
  \dodoi{10.1111/j.1365-2966.2012.21373.x}

\bibitem[{{Hankins} {et~al.}(2019){Hankins}, {Lau}, {Mills}, {Morris}, \&
  {Herter}}]{hankins2019}
{Hankins}, M.~J., {Lau}, R.~M., {Mills}, E.~A.~C., {Morris}, M.~R., \&
  {Herter}, T.~L. 2019, \apj, 877, 22, \dodoi{10.3847/1538-4357/ab174e}

\bibitem[{{Hankins} {et~al.}(2017){Hankins}, {Lau}, {Morris}, \&
  {Herter}}]{hankins2017}
{Hankins}, M.~J., {Lau}, R.~M., {Morris}, M.~R., \& {Herter}, T.~L. 2017, \apj,
  837, 79, \dodoi{10.3847/1538-4357/aa5f5b}

\bibitem[{{Henshaw} {et~al.}(2019){Henshaw}, {Ginsburg}, {Haworth}, {Longmore},
  {Kruijssen}, {Mills}, {Sokolov}, {Walker}, {Barnes}, {Contreras}, {Bally},
  {Battersby}, {Beuther}, {Butterfield}, {Dale}, {Henning}, {Jackson},
  {Kauffmann}, {Pillai}, {Ragan}, {Riener}, \& {Zhang}}]{henshaw2019}
{Henshaw}, J.~D., {Ginsburg}, A., {Haworth}, T.~J., {et~al.} 2019, \mnras, 485,
  2457, \dodoi{10.1093/mnras/stz471}

\bibitem[{{Houghton} \& {Whiteoak}(1995)}]{houghton1995}
{Houghton}, S., \& {Whiteoak}, J.~B. 1995, \mnras, 273, 1033,
  \dodoi{10.1093/mnras/273.4.1033}

\bibitem[{{Immer} {et~al.}(2012{\natexlab{a}}){Immer}, {Menten}, {Schuller}, \&
  {Lis}}]{immer2012b}
{Immer}, K., {Menten}, K.~M., {Schuller}, F., \& {Lis}, D.~C.
  2012{\natexlab{a}}, \aap, 548, A120, \dodoi{10.1051/0004-6361/201219182}

\bibitem[{{Immer} {et~al.}(2012{\natexlab{b}}){Immer}, {Schuller}, {Omont}, \&
  {Menten}}]{immer2012a}
{Immer}, K., {Schuller}, F., {Omont}, A., \& {Menten}, K.~M.
  2012{\natexlab{b}}, \aap, 537, A121, \dodoi{10.1051/0004-6361/201117857}

\bibitem[{{Jeffreson} {et~al.}(2018){Jeffreson}, {Kruijssen}, {Krumholz}, \&
  {Longmore}}]{jeffreson2018}
{Jeffreson}, S.~M.~R., {Kruijssen}, J.~M.~D., {Krumholz}, M.~R., \& {Longmore},
  S.~N. 2018, \mnras, 478, 3380, \dodoi{10.1093/mnras/sty1154}

\bibitem[{{Jones} {et~al.}(2013){Jones}, {Burton}, {Cunningham}, {Tothill}, \&
  {Walsh}}]{jones2013}
{Jones}, P.~A., {Burton}, M.~G., {Cunningham}, M.~R., {Tothill}, N.~F.~H., \&
  {Walsh}, A.~J. 2013, \mnras, 433, 221, \dodoi{10.1093/mnras/stt717}

\bibitem[{{Kauffmann} \& {Pillai}(2010)}]{kauffmann2010}
{Kauffmann}, J., \& {Pillai}, T. 2010, \apjl, 723, L7,
  \dodoi{10.1088/2041-8205/723/1/L7}

\bibitem[{{Kauffmann} {et~al.}(2013){Kauffmann}, {Pillai}, \&
  {Zhang}}]{kauffmann2013a}
{Kauffmann}, J., {Pillai}, T., \& {Zhang}, Q. 2013, \apjl, 765, L35,
  \dodoi{10.1088/2041-8205/765/2/L35}

\bibitem[{{Kauffmann} {et~al.}(2017{\natexlab{a}}){Kauffmann}, {Pillai},
  {Zhang}, {Menten}, {Goldsmith}, {Lu}, \& {Guzm{\'a}n}}]{kauffmann2017a}
{Kauffmann}, J., {Pillai}, T., {Zhang}, Q., {et~al.} 2017{\natexlab{a}}, \aap,
  603, A89, \dodoi{10.1051/0004-6361/201628088}

\bibitem[{{Kauffmann} {et~al.}(2017{\natexlab{b}}){Kauffmann}, {Pillai},
  {Zhang}, {Menten}, {Goldsmith}, {Lu}, {Guzm{\'a}n}, \&
  {Schmiedeke}}]{kauffmann2017b}
---. 2017{\natexlab{b}}, \aap, 603, A90, \dodoi{10.1051/0004-6361/201628089}

\bibitem[{{Kendrew} {et~al.}(2013){Kendrew}, {Ginsburg}, {Johnston}, {Beuther},
  {Bally}, {Cyganowski}, \& {Battersby}}]{kendrew2013}
{Kendrew}, S., {Ginsburg}, A., {Johnston}, K., {et~al.} 2013, \apjl, 775, L50,
  \dodoi{10.1088/2041-8205/775/2/L50}

\bibitem[{{Koepferl} {et~al.}(2015){Koepferl}, {Robitaille}, {Morales}, \&
  {Johnston}}]{koepferl2015}
{Koepferl}, C.~M., {Robitaille}, T.~P., {Morales}, E. F.~E., \& {Johnston},
  K.~G. 2015, \apj, 799, 53, \dodoi{10.1088/0004-637X/799/1/53}

\bibitem[{{Krieger} {et~al.}(2017){Krieger}, {Ott}, {Beuther}, {Walter},
  {Kruijssen}, {Meier}, {Mills}, {Contreras}, {Edwards}, {Ginsburg}, {Henkel},
  {Henshaw}, {Jackson}, {Kauffmann}, {Longmore}, {Mart{\'{\i}}n}, {Morris},
  {Pillai}, {Rickert}, {Rosolowsky}, {Shinnaga}, {Walsh}, {Yusef-Zadeh}, \&
  {Zhang}}]{krieger2017}
{Krieger}, N., {Ott}, J., {Beuther}, H., {et~al.} 2017, \apj, 850, 77,
  \dodoi{10.3847/1538-4357/aa951c}

\bibitem[{{Kroupa}(2001)}]{kroupa2001}
{Kroupa}, P. 2001, \mnras, 322, 231, \dodoi{10.1046/j.1365-8711.2001.04022.x}

\bibitem[{{Kruijssen} {et~al.}(2015){Kruijssen}, {Dale}, \&
  {Longmore}}]{kruijssen2015}
{Kruijssen}, J.~M.~D., {Dale}, J.~E., \& {Longmore}, S.~N. 2015, \mnras, 447,
  1059, \dodoi{10.1093/mnras/stu2526}

\bibitem[{{Kruijssen} {et~al.}(2014){Kruijssen}, {Longmore}, {Elmegreen},
  {Murray}, {Bally}, {Testi}, \& {Kennicutt}}]{kruijssen2014}
{Kruijssen}, J.~M.~D., {Longmore}, S.~N., {Elmegreen}, B.~G., {et~al.} 2014,
  \mnras, 440, 3370, \dodoi{10.1093/mnras/stu494}

\bibitem[{{Kruijssen} {et~al.}(2019){Kruijssen}, {Dale}, {Longmore}, {Walker},
  {Henshaw}, {Jeffreson}, {Petkova}, {Ginsburg}, {Barnes}, {Battersby},
  {Immer}, {Jackson}, {Keto}, {Krieger}, {Mills}, {S{\'a}nchez-Monge},
  {Schmiedeke}, {Suri}, \& {Zhang}}]{kruijssen2019}
{Kruijssen}, J.~M.~D., {Dale}, J.~E., {Longmore}, S.~N., {et~al.} 2019, \mnras,
  484, 5734, \dodoi{10.1093/mnras/stz381}

\bibitem[{{Krumholz} \& {Kruijssen}(2015)}]{KK2015}
{Krumholz}, M.~R., \& {Kruijssen}, J.~M.~D. 2015, \mnras, 453, 739,
  \dodoi{10.1093/mnras/stv1670}

\bibitem[{{Krumholz} {et~al.}(2017){Krumholz}, {Kruijssen}, \&
  {Crocker}}]{krumholz2017}
{Krumholz}, M.~R., {Kruijssen}, J.~M.~D., \& {Crocker}, R.~M. 2017, \mnras,
  466, 1213, \dodoi{10.1093/mnras/stw3195}

\bibitem[{{Krumholz} \& {McKee}(2008)}]{krumholz2008}
{Krumholz}, M.~R., \& {McKee}, C.~F. 2008, \nat, 451, 1082,
  \dodoi{10.1038/nature06620}

\bibitem[{{Lada} {et~al.}(2010){Lada}, {Lombardi}, \& {Alves}}]{lada2010}
{Lada}, C.~J., {Lombardi}, M., \& {Alves}, J.~F. 2010, \apj, 724, 687,
  \dodoi{10.1088/0004-637X/724/1/687}

\bibitem[{{Lang} {et~al.}(2010){Lang}, {Goss}, {Cyganowski}, \&
  {Clubb}}]{lang2010}
{Lang}, C.~C., {Goss}, W.~M., {Cyganowski}, C., \& {Clubb}, K.~I. 2010, \apjs,
  191, 275, \dodoi{10.1088/0067-0049/191/2/275}

\bibitem[{{Lang} {et~al.}(2001){Lang}, {Goss}, \& {Morris}}]{lang2001}
{Lang}, C.~C., {Goss}, W.~M., \& {Morris}, M. 2001, \aj, 121, 2681,
  \dodoi{10.1086/320373}

\bibitem[{{Lang} {et~al.}(2005){Lang}, {Johnson}, {Goss}, \&
  {Rodr{\'\i}guez}}]{lang2005}
{Lang}, C.~C., {Johnson}, K.~E., {Goss}, W.~M., \& {Rodr{\'\i}guez}, L.~F.
  2005, \aj, 130, 2185, \dodoi{10.1086/496976}

\bibitem[{{Lazio} \& {Cordes}(2008)}]{lazio2008}
{Lazio}, T. J.~W., \& {Cordes}, J.~M. 2008, \apjs, 174, 481,
  \dodoi{10.1086/521676}

\bibitem[{{Liszt} \& {Spiker}(1995)}]{liszt1995}
{Liszt}, H.~S., \& {Spiker}, R.~W. 1995, \apjs, 98, 259, \dodoi{10.1086/192160}

\bibitem[{{Lo} \& {Claussen}(1983)}]{lo1983}
{Lo}, K.~Y., \& {Claussen}, M.~J. 1983, \nat, 306, 647,
  \dodoi{10.1038/306647a0}

\bibitem[{{Longmore} {et~al.}(2013{\natexlab{a}}){Longmore}, {Bally}, {Testi},
  {Purcell}, {Walsh}, {Bressert}, {Pestalozzi}, {Molinari}, {Ott}, {Cortese},
  {Battersby}, {Murray}, {Lee}, {Kruijssen}, {Schisano}, \&
  {Elia}}]{longmore2013a}
{Longmore}, S.~N., {Bally}, J., {Testi}, L., {et~al.} 2013{\natexlab{a}},
  \mnras, 429, 987, \dodoi{10.1093/mnras/sts376}

\bibitem[{{Longmore} {et~al.}(2013{\natexlab{b}}){Longmore}, {Kruijssen},
  {Bally}, {Ott}, {Testi}, {Rathborne}, {Bastian}, {Bressert}, {Molinari},
  {Battersby}, \& {Walsh}}]{longmore2013b}
{Longmore}, S.~N., {Kruijssen}, J.~M.~D., {Bally}, J., {et~al.}
  2013{\natexlab{b}}, \mnras, 433, L15, \dodoi{10.1093/mnrasl/slt048}

\bibitem[{{Lu} {et~al.}(2015){Lu}, {Zhang}, {Kauffmann}, {Pillai}, {Longmore},
  {Kruijssen}, {Battersby}, \& {Gu}}]{lu2015b}
{Lu}, X., {Zhang}, Q., {Kauffmann}, J., {et~al.} 2015, \apjl, 814, L18,
  \dodoi{10.1088/2041-8205/814/2/L18}

\bibitem[{{Lu} {et~al.}(2017){Lu}, {Zhang}, {Kauffmann}, {Pillai}, {Longmore},
  {Kruijssen}, {Battersby}, {Liu}, {Ginsburg}, {Mills}, {Zhang}, \&
  {Gu}}]{lu2017}
---. 2017, \apj, 839, 1, \dodoi{10.3847/1538-4357/aa67f7}

\bibitem[{{Lu} {et~al.}(2019){Lu}, {Zhang}, {Kauffmann}, {Pillai}, {Ginsburg},
  {Mills}, {Kruijssen}, {Longmore}, {Battersby}, {Liu}, \& {Gu}}]{lu2019a}
---. 2019, \apj, 872, 171, \dodoi{10.3847/1538-4357/ab017d}

\bibitem[{{Ludovici} {et~al.}(2016){Ludovici}, {Lang}, {Morris}, {Mutel},
  {Mills}, {Toomey}, \& {Ott}}]{ludovici2016}
{Ludovici}, D.~A., {Lang}, C.~C., {Morris}, M.~R., {et~al.} 2016, \apj, 826,
  218, \dodoi{10.3847/0004-637X/826/2/218}

\bibitem[{{Mauerhan} {et~al.}(2010{\natexlab{a}}){Mauerhan}, {Cotera}, {Dong},
  {Morris}, {Wang}, {Stolovy}, \& {Lang}}]{mauerhan2010b}
{Mauerhan}, J.~C., {Cotera}, A., {Dong}, H., {et~al.} 2010{\natexlab{a}}, \apj,
  725, 188, \dodoi{10.1088/0004-637X/725/1/188}

\bibitem[{{Mauerhan} {et~al.}(2010{\natexlab{b}}){Mauerhan}, {Muno}, {Morris},
  {Stolovy}, \& {Cotera}}]{mauerhan2010a}
{Mauerhan}, J.~C., {Muno}, M.~P., {Morris}, M.~R., {Stolovy}, S.~R., \&
  {Cotera}, A. 2010{\natexlab{b}}, \apj, 710, 706,
  \dodoi{10.1088/0004-637X/710/1/706}

\bibitem[{{McMullin} {et~al.}(2007){McMullin}, {Waters}, {Schiebel}, {Young},
  \& {Golap}}]{mcmullin2007}
{McMullin}, J.~P., {Waters}, B., {Schiebel}, D., {Young}, W., \& {Golap}, K.
  2007, in ASP Conf.~Ser., Vol. 376, Astronomical Data Analysis Software and
  Systems XVI, ed. R.~A. {Shaw}, F.~{Hill}, \& D.~J. {Bell {(San Francisco, CA:
  ASP)}}, 127

\bibitem[{{Mehringer} {et~al.}(1994){Mehringer}, {Goss}, \&
  {Palmer}}]{mehringer1994}
{Mehringer}, D.~M., {Goss}, W.~M., \& {Palmer}, P. 1994, \apj, 434, 237,
  \dodoi{10.1086/174721}

\bibitem[{{Mehringer} {et~al.}(1995){Mehringer}, {Palmer}, \&
  {Goss}}]{mehringer1995}
{Mehringer}, D.~M., {Palmer}, P., \& {Goss}, W.~M. 1995, \apjs, 97, 497,
  \dodoi{10.1086/192148}

\bibitem[{{Mehringer} {et~al.}(1993){Mehringer}, {Palmer}, {Goss}, \&
  {Yusef-Zadeh}}]{mehringer1993}
{Mehringer}, D.~M., {Palmer}, P., {Goss}, W.~M., \& {Yusef-Zadeh}, F. 1993,
  \apj, 412, 684, \dodoi{10.1086/172954}

\bibitem[{{Meidt} {et~al.}(2018){Meidt}, {Leroy}, {Rosolowsky}, {Kruijssen},
  {Schinnerer}, {Schruba}, {Pety}, {Blanc}, {Bigiel}, {Chevance}, {Hughes},
  {Querejeta}, \& {Usero}}]{meidt2018}
{Meidt}, S.~E., {Leroy}, A.~K., {Rosolowsky}, E., {et~al.} 2018, \apj, 854,
  100, \dodoi{10.3847/1538-4357/aaa290}

\bibitem[{{Menten}(1991)}]{menten1991a}
{Menten}, K. 1991, in ASP Conf.\ Ser., Vol.~16, Atoms, Ions and Molecules: New
  Results in Spectral Line Astrophysics, ed. A.~D. {Haschick} \& P.~T.~P. {Ho
  (San Francisco: ASP)}, 119

\bibitem[{{Mezger} {et~al.}(1974){Mezger}, {Smith}, \&
  {Churchwell}}]{mezger1974}
{Mezger}, P.~G., {Smith}, L.~F., \& {Churchwell}, E. 1974, \aap, 32, 269

\bibitem[{{Mills} {et~al.}(2011){Mills}, {Morris}, {Lang}, {Dong}, {Wang},
  {Cotera}, \& {Stolovy}}]{mills2011}
{Mills}, E., {Morris}, M.~R., {Lang}, C.~C., {et~al.} 2011, \apj, 735, 84,
  \dodoi{10.1088/0004-637X/735/2/84}

\bibitem[{{Mills} {et~al.}(2015){Mills}, {Butterfield}, {Ludovici}, {Lang},
  {Ott}, {Morris}, \& {Schmitz}}]{mills2015}
{Mills}, E.~A.~C., {Butterfield}, N., {Ludovici}, D.~A., {et~al.} 2015, \apj,
  805, 72, \dodoi{10.1088/0004-637X/805/1/72}

\bibitem[{{Molinari} {et~al.}(2010){Molinari}, {Swinyard}, {Bally}, {Barlow},
  {Bernard}, {Martin}, {Moore}, {Noriega-Crespo}, {Plume}, {Testi}, {Zavagno},
  {Abergel}, {Ali}, {Anderson}, {Andr{\'e}}, {Baluteau}, {Battersby},
  {Beltr{\'a}n}, {Benedettini}, {Billot}, {Blommaert}, {Bontemps}, {Boulanger},
  {Brand}, {Brunt}, {Burton}, {Calzoletti}, {Carey}, {Caselli}, {Cesaroni},
  {Cernicharo}, {Chakrabarti}, {Chrysostomou}, {Cohen}, {Compiegne}, {de
  Bernardis}, {de Gasperis}, {di Giorgio}, {Elia}, {Faustini}, {Flagey},
  {Fukui}, {Fuller}, {Ganga}, {Garcia-Lario}, {Glenn}, {Goldsmith}, {Griffin},
  {Hoare}, {Huang}, {Ikhenaode}, {Joblin}, {Joncas}, {Juvela}, {Kirk},
  {Lagache}, {Li}, {Lim}, {Lord}, {Marengo}, {Marshall}, {Masi}, {Massi},
  {Matsuura}, {Minier}, {Miville-Desch{\^e}nes}, {Montier}, {Morgan}, {Motte},
  {Mottram}, {M{\"u}ller}, {Natoli}, {Neves}, {Olmi}, {Paladini}, {Paradis},
  {Parsons}, {Peretto}, {Pestalozzi}, {Pezzuto}, {Piacentini}, {Piazzo},
  {Polychroni}, {Pomar{\`e}s}, {Popescu}, {Reach}, {Ristorcelli}, {Robitaille},
  {Robitaille}, {Rod{\'o}n}, {Roy}, {Royer}, {Russeil}, {Saraceno}, {Sauvage},
  {Schilke}, {Schisano}, {Schneider}, {Schuller}, {Schulz}, {Sibthorpe},
  {Smith}, {Smith}, {Spinoglio}, {Stamatellos}, {Strafella}, {Stringfellow},
  {Sturm}, {Taylor}, {Thompson}, {Traficante}, {Tuffs}, {Umana}, {Valenziano},
  {Vavrek}, {Veneziani}, {Viti}, {Waelkens}, {Ward-Thompson}, {White},
  {Wilcock}, {Wyrowski}, {Yorke}, \& {Zhang}}]{molinari2010}
{Molinari}, S., {Swinyard}, B., {Bally}, J., {et~al.} 2010, \aap, 518, L100,
  \dodoi{10.1051/0004-6361/201014659}

\bibitem[{{Muno} {et~al.}(2006){Muno}, {Bauer}, {Bandyopadhyay}, \&
  {Wang}}]{muno2006}
{Muno}, M.~P., {Bauer}, F.~E., {Bandyopadhyay}, R.~M., \& {Wang}, Q.~D. 2006,
  \apjs, 165, 173, \dodoi{10.1086/504798}

\bibitem[{{Muno} {et~al.}(2009){Muno}, {Bauer}, {Baganoff}, {Band yopadhyay},
  {Bower}, {Brandt}, {Broos}, {Cotera}, {Eikenberry}, {Garmire}, {Hyman},
  {Kassim}, {Lang}, {Lazio}, {Law}, {Mauerhan}, {Morris}, {Nagata},
  {Nishiyama}, {Park}, {Ram{\`\i}rez}, {Stolovy}, {Wijnands}, {Wang}, {Wang},
  \& {Yusef-Zadeh}}]{muno2009}
{Muno}, M.~P., {Bauer}, F.~E., {Baganoff}, F.~K., {et~al.} 2009, \apjs, 181,
  110, \dodoi{10.1088/0067-0049/181/1/110}

\bibitem[{{Parker} {et~al.}(2016){Parker}, {Boji{\v{c}}i{\'c}}, \&
  {Frew}}]{parker2016}
{Parker}, Q.~A., {Boji{\v{c}}i{\'c}}, I.~S., \& {Frew}, D.~J. 2016, JPhCS, 728,
  032008, \dodoi{10.1088/1742-6596/728/3/032008}

\bibitem[{{Rathborne} {et~al.}(2015){Rathborne}, {Longmore}, {Jackson},
  {Alves}, {Bally}, {Bastian}, {Contreras}, {Foster}, {Garay}, {Kruijssen},
  {Testi}, \& {Walsh}}]{rathborne2015}
{Rathborne}, J.~M., {Longmore}, S.~N., {Jackson}, J.~M., {et~al.} 2015, \apj,
  802, 125, \dodoi{10.1088/0004-637X/802/2/125}

\bibitem[{{Rickert} {et~al.}(2019){Rickert}, {Yusef-Zadeh}, \&
  {Ott}}]{rickert2019}
{Rickert}, M., {Yusef-Zadeh}, F., \& {Ott}, J. 2019, \mnras, 482, 5349,
  \dodoi{10.1093/mnras/sty2901}

\bibitem[{{Roberts} \& {Goss}(1993)}]{roberts1993}
{Roberts}, D.~A., \& {Goss}, W.~M. 1993, \apjs, 86, 133, \dodoi{10.1086/191773}

\bibitem[{{Robitaille} \& {Bressert}(2012)}]{aplpy2012}
{Robitaille}, T., \& {Bressert}, E. 2012, {APLpy: Astronomical Plotting Library
  in Python},  Astrophysics Source Code Library.
\newblock \doeprint{1208.017}

\bibitem[{{Rodr{\'{\i}}guez} \& {Zapata}(2013)}]{rodriguez2013}
{Rodr{\'{\i}}guez}, L.~F., \& {Zapata}, L.~A. 2013, \apjl, 767, L13,
  \dodoi{10.1088/2041-8205/767/1/L13}

\bibitem[{{S{\'a}nchez-Monge} {et~al.}(2013){S{\'a}nchez-Monge}, {Kurtz},
  {Palau}, {Estalella}, {Shepherd}, {Lizano}, {Franco}, \&
  {Garay}}]{sanchezmonge2013a}
{S{\'a}nchez-Monge}, {\'A}., {Kurtz}, S., {Palau}, A., {et~al.} 2013, \apj,
  766, 114, \dodoi{10.1088/0004-637X/766/2/114}

\bibitem[{{Sewilo} {et~al.}(2004){Sewilo}, {Churchwell}, {Kurtz}, {Goss}, \&
  {Hofner}}]{sewilo2004}
{Sewilo}, M., {Churchwell}, E., {Kurtz}, S., {Goss}, W.~M., \& {Hofner}, P.
  2004, \apj, 605, 285, \dodoi{10.1086/382268}

\bibitem[{{Tsuboi} {et~al.}(2016){Tsuboi}, {Kitamura}, {Miyoshi}, {Uehara},
  {Tsutsumi}, \& {Miyazaki}}]{tsuboi2016}
{Tsuboi}, M., {Kitamura}, Y., {Miyoshi}, M., {et~al.} 2016, \pasj, 68, L7,
  \dodoi{10.1093/pasj/psw031}

\bibitem[{{Tsuboi} {et~al.}(2017){Tsuboi}, {Kitamura}, {Uehara}, {Miyawaki},
  {Tsutsumi}, {Miyazaki}, \& {Miyoshi}}]{tsuboi2017}
{Tsuboi}, M., {Kitamura}, Y., {Uehara}, K., {et~al.} 2017, \apj, 842, 94,
  \dodoi{10.3847/1538-4357/aa74e3}

\bibitem[{{Tsujimoto} {et~al.}(2006){Tsujimoto}, {Hosokawa}, {Feigelson},
  {Getman}, \& {Broos}}]{tsujimoto2006}
{Tsujimoto}, M., {Hosokawa}, T., {Feigelson}, E.~D., {Getman}, K.~V., \&
  {Broos}, P.~S. 2006, \apj, 653, 409, \dodoi{10.1086/507439}

\bibitem[{{van der Walt}(2014)}]{vanderwalt2014}
{van der Walt}, D.~J. 2014, \aap, 562, A68, \dodoi{10.1051/0004-6361/201322512}

\bibitem[{{Walker} {et~al.}(2015){Walker}, {Longmore}, {Bastian}, {Kruijssen},
  {Rathborne}, {Jackson}, {Foster}, \& {Contreras}}]{walker2015}
{Walker}, D.~L., {Longmore}, S.~N., {Bastian}, N., {et~al.} 2015, \mnras, 449,
  715, \dodoi{10.1093/mnras/stv300}

\bibitem[{{Walker} {et~al.}(2018){Walker}, {Longmore}, {Zhang}, {Battersby},
  {Keto}, {Kruijssen}, {Ginsburg}, {Lu}, {Henshaw}, {Kauffmann}, {Pillai},
  {Mills}, {Walsh}, {Bally}, {Ho}, {Immer}, \& {Johnston}}]{walker2018}
{Walker}, D.~L., {Longmore}, S.~N., {Zhang}, Q., {et~al.} 2018, \mnras, 474,
  2373, \dodoi{10.1093/mnras/stx2898}

\bibitem[{{Walsh} {et~al.}(2011){Walsh}, {Breen}, {Britton}, {Brooks},
  {Burton}, {Cunningham}, {Green}, {Harvey-Smith}, {Hindson}, {Hoare},
  {Indermuehle}, {Jones}, {Lo}, {Longmore}, {Lowe}, {Phillips}, {Purcell},
  {Thompson}, {Urquhart}, {Voronkov}, {White}, \& {Whiting}}]{walsh2011}
{Walsh}, A.~J., {Breen}, S.~L., {Britton}, T., {et~al.} 2011, \mnras, 416,
  1764, \dodoi{10.1111/j.1365-2966.2011.19115.x}

\bibitem[{{Wang} {et~al.}(2010){Wang}, {Dong}, {Cotera}, {Stolovy}, {Morris},
  {Lang}, {Muno}, {Schneider}, \& {Calzetti}}]{wang2010}
{Wang}, Q.~D., {Dong}, H., {Cotera}, A., {et~al.} 2010, \mnras, 402, 895,
  \dodoi{10.1111/j.1365-2966.2009.15973.x}

\bibitem[{{Wang} {et~al.}(2018){Wang}, {Bihr}, {Rugel}, {Beuther}, {Johnston},
  {Ott}, {Soler}, {Brunthaler}, {Anderson}, {Urquhart}, {Klessen}, {Linz},
  {McClure-Griffiths}, {Glover}, {Menten}, {Bigiel}, {Hoare}, \&
  {Longmore}}]{wang2018}
{Wang}, Y., {Bihr}, S., {Rugel}, M., {et~al.} 2018, \aap, 619, A124,
  \dodoi{10.1051/0004-6361/201833642}

\bibitem[{{Xu} {et~al.}(2008){Xu}, {Li}, {Hachisuka}, {Pandian}, {Menten}, \&
  {Henkel}}]{xu2008}
{Xu}, Y., {Li}, J.~J., {Hachisuka}, K., {et~al.} 2008, \aap, 485, 729,
  \dodoi{10.1051/0004-6361:200809472}

\bibitem[{{Yusef-Zadeh} {et~al.}(2004){Yusef-Zadeh}, {Hewitt}, \&
  {Cotton}}]{yusefzadeh2004}
{Yusef-Zadeh}, F., {Hewitt}, J.~W., \& {Cotton}, W. 2004, \apjs, 155, 421,
  \dodoi{10.1086/425257}

\bibitem[{{Yusef-Zadeh} {et~al.}(2010){Yusef-Zadeh}, {Lacy}, {Wardle},
  {Whitney}, {Bushouse}, {Roberts}, \& {Arendt}}]{yusefzadeh2010}
{Yusef-Zadeh}, F., {Lacy}, J.~H., {Wardle}, M., {et~al.} 2010, \apj, 725, 1429,
  \dodoi{10.1088/0004-637X/725/2/1429}

\bibitem[{{Yusef-Zadeh} \& {Morris}(1987{\natexlab{a}})}]{yusefzadeh1987a}
{Yusef-Zadeh}, F., \& {Morris}, M. 1987{\natexlab{a}}, \aj, 94, 1178,
  \dodoi{10.1086/114555}

\bibitem[{{Yusef-Zadeh} \& {Morris}(1987{\natexlab{b}})}]{yusefzadeh1987b}
---. 1987{\natexlab{b}}, \apj, 320, 545, \dodoi{10.1086/165572}

\bibitem[{{Yusef-Zadeh} {et~al.}(1984){Yusef-Zadeh}, {Morris}, \&
  {Chance}}]{yusefzadeh1984}
{Yusef-Zadeh}, F., {Morris}, M., \& {Chance}, D. 1984, \nat, 310, 557,
  \dodoi{10.1038/310557a0}

\bibitem[{{Yusef-Zadeh} {et~al.}(2009){Yusef-Zadeh}, {Hewitt}, {Arendt},
  {Whitney}, {Rieke}, {Wardle}, {Hinz}, {Stolovy}, {Lang}, {Burton}, \&
  {Ramirez}}]{yusefzadeh2009}
{Yusef-Zadeh}, F., {Hewitt}, J.~W., {Arendt}, R.~G., {et~al.} 2009, \apj, 702,
  178, \dodoi{10.1088/0004-637X/702/1/178}

\bibitem[{{Zhao} {et~al.}(2013){Zhao}, {Morris}, \& {Goss}}]{zhao2013}
{Zhao}, J.-H., {Morris}, M.~R., \& {Goss}, W.~M. 2013, \apj, 777, 146,
  \dodoi{10.1088/0004-637X/777/2/146}

\bibitem[{{Zhao} {et~al.}(2016){Zhao}, {Morris}, \& {Goss}}]{zhao2016}
---. 2016, \apj, 817, 171, \dodoi{10.3847/0004-637X/817/2/171}

\bibitem[{{Zhao} {et~al.}(2009){Zhao}, {Morris}, {Goss}, \& {An}}]{zhao2009}
{Zhao}, J.-H., {Morris}, M.~R., {Goss}, W.~M., \& {An}, T. 2009, \apj, 699,
  186, \dodoi{10.1088/0004-637X/699/1/186}

\bibitem[{{Zhao} \& {Wright}(2011)}]{zhao2011}
{Zhao}, J.-H., \& {Wright}, M.~C.~H. 2011, \apj, 742, 50,
  \dodoi{10.1088/0004-637X/742/1/50}

\bibitem[{{Zhu} {et~al.}(2018){Zhu}, {Li}, \& {Morris}}]{zhu2018}
{Zhu}, Z., {Li}, Z., \& {Morris}, M.~R. 2018, \apjs, 235, 26,
  \dodoi{10.3847/1538-4365/aab14f}

\bibitem[{{Zoonematkermani} {et~al.}(1990){Zoonematkermani}, {Helfand},
  {Becker}, {White}, \& {Perley}}]{zoonematkermani1990}
{Zoonematkermani}, S., {Helfand}, D.~J., {Becker}, R.~H., {White}, R.~L., \&
  {Perley}, R.~A. 1990, \apjs, 74, 181, \dodoi{10.1086/191496}

\end{thebibliography}
%\bibliography{my}

\clearpage

\movetabledown=15mm
\begin{longrotatetable}
\begin{deluxetable*}{ccccccccccccc}
\tabletypesize{\scriptsize}
\tablecaption{Properties of Compact Sources in $C$-band Continuum.\label{tab:uchii}}
\tablewidth{0pt}
\tablehead{
\colhead{ID} & R.A. \& Decl.  & $r_\text{eff}$\tablenotemark{a} & $I_\text{peak}$\tablenotemark{b} & $F_\text{int}$\tablenotemark{b} & $\alpha\pm\sigma(\alpha)$\tablenotemark{b,c} & HST Pa-$\alpha$ & WISE \hii{} Regions &  \textit{Spitzer} YSOs & \textit{Herschel} $N$(H$_2$) & Counterparts\tablenotemark{d} & Previous Detections\tablenotemark{e} &  Remarks \\
 & (J2000) & (pc) & (\mjypbm{}) & (mJy) &  & (Ref: W10) & (Ref: A14) & (Refs: Y09, A11) & (Ref: B11) & Against UC~\hii{}  & of Compact Continuum Sources&
}
\startdata
%%%%
C1   & 17:46:53.82, $-$28:19:00.38 & 0.01 & 0.99$\pm$0.19 & 1.19$\pm$0.19 & $-$2.67$\pm$0.26 & \nodata & G000.674$+$00.083 & & N & X(M06) & &  \\
C2   & 17:47:40.33, $-$28:20:06.11 & 0.03 & 2.95$\pm$0.23 & 3.25$\pm$0.23 & $-$0.14$\pm$0.01 & \nodata & & & Y & & & UC? \\
C3   & 17:47:43.10, $-$28:20:30.38 & 0.02 & 1.99$\pm$0.19 & 1.69$\pm$0.17 & $-$0.24$\pm$0.01 & \nodata & & & Y & & & \\
C4   & 17:47:22.61, $-$28:24:54.22 & UR   & 2.70$\pm$0.54 & 3.11$\pm$0.54 & $-$2.71$\pm$1.81 & \nodata & & & Y  & & & \\
C5   & 17:47:23.36, $-$28:25:33.81 & 0.02 & 3.23$\pm$0.58 & 4.57$\pm$0.60 & $-$1.61$\pm$0.94 & \nodata & & & Y & X(M06,M09) & &  \\
C6   & 17:46:38.51, $-$28:25:34.95 & 0.06 & 7.98$\pm$0.46 & 12.09$\pm$0.63& $-$0.48$\pm$0.01 & \nodata & & & N & & & \\
C7   & 17:46:37.94, $-$28:25:51.14 & 0.03 & 2.60$\pm$0.25 & 2.98$\pm$0.26 &    0.82$\pm$0.03 & \nodata & & & N & & & \\
C8   & 17:46:37.57, $-$28:26:00.44 & 0.03 & 2.71$\pm$0.28 & 3.20$\pm$0.28 & $-$0.52$\pm$0.02 & \nodata & & & N & & & \\
C9   & 17:46:55.61, $-$28:26:07.60 & 0.04 & 1.35$\pm$0.19 & 3.56$\pm$0.25 &    0.43$\pm$0.02 & \nodata & & & Y & & & UC? \\
C10  & 17:47:27.09, $-$28:27:24.80 & 0.03 & 1.45$\pm$0.26 & 3.62$\pm$0.30 & $-$2.32$\pm$1.86 & \nodata & & SSTGC805200 & Y & & & UC? \\
C11  & 17:46:41.77, $-$28:28:16.71 & 0.02 & 3.53$\pm$0.30 & 3.19$\pm$0.29 & $-$1.10$\pm$0.03 & \nodata & & & Y & & 2LC000.520$+$0.040(L08) & \\
C12  & 17:46:41.86, $-$28:28:18.21 & 0.02 & 2.14$\pm$0.27 & 2.10$\pm$0.26 & $-$3.50$\pm$0.63 & \nodata & & & Y & & & \\
C13  & 17:46:54.52, $-$28:31:16.21 & 0.04 & 3.87$\pm$0.27 & 3.79$\pm$0.25 &    0.00$\pm$0.01 & \nodata & & & N & & & \\
C14  & 17:46:30.17, $-$28:32:13.62 & 0.02 & 0.42$\pm$0.06 & 0.45$\pm$0.06 & $-$0.70$\pm$0.57 & \nodata & & & Y & & & UC? \\
C15  & 17:46:21.38, $-$28:32:27.39 & 0.04 & 0.45$\pm$0.07 & 0.85$\pm$0.08 & $-$1.60$\pm$0.98 & \nodata & & & Y & & 2LC000.422$+$0.068(L08) &  \\
C16  & 17:46:54.28, $-$28:32:39.01 & 0.13 & 6.96$\pm$0.41 & 25.60$\pm$1.29& $-$0.04$\pm$0.01 & \nodata & & & N & & 2LC000.481$-$0.037(L08) &\\
C17  & 17:46:45.90, $-$28:32:55.88 & 0.03 & 0.99$\pm$0.12 & 0.97$\pm$0.12 & $-$1.69$\pm$0.59 & \nodata & & & Y & & & \\
C18  & 17:46:31.14, $-$28:33:18.42 & 0.03 & 0.43$\pm$0.07 & 0.74$\pm$0.08 & $-$2.52$\pm$1.87 & \nodata & & & Y & & & \\
C19  & 17:46:52.09, $-$28:33:31.53 & 0.03 & 0.72$\pm$0.13 & 1.15$\pm$0.13 & $-$0.24$\pm$0.11 & \nodata & & & N & & & \\
C20  & 17:47:08.34, $-$28:34:47.82 & 0.10 & 6.77$\pm$0.41 & 15.14$\pm$0.78& $-$0.80$\pm$0.05 & \nodata & & & N & & 2LC000.477-0.100(L08) & \\
C21  & 17:47:07.02, $-$28:34:58.94 & 0.02 & 0.86$\pm$0.15 & 1.16$\pm$0.16 & $-$1.72$\pm$0.70 & \nodata & & & N & & & \\
C22  & 17:47:06.06, $-$28:35:10.36 & 0.07 & 4.14$\pm$0.27 & 6.48$\pm$0.36 & $-$1.24$\pm$0.11 & \nodata & & & N & & & \\
C23  & 17:46:21.42, $-$28:35:38.49 & 0.03 & 0.90$\pm$0.07 & 0.83$\pm$0.07 & $-$0.22$\pm$0.10 & \nodata & & SSTGC639320, EGO g16 & Y & & & UC \\
C24  & 17:46:40.98, $-$28:35:40.90 & 0.08 & 3.18$\pm$0.21 & 6.77$\pm$0.36 & $-$2.27$\pm$0.23 & \nodata & & & N & & 2LC000.413-0.022(L08) &  \\
C25  & 17:46:51.33, $-$28:36:09.94 & 0.09 & 7.06$\pm$0.40 & 10.61$\pm$0.55& $-$2.32$\pm$0.11 & \nodata & & & N & X(M06) & 2LC000.426-0.058(L08)&  \\
C26  & 17:46:22.01, $-$28:36:13.00 & 0.06 & 0.32$\pm$0.06 & 1.29$\pm$0.09 & $-$1.81$\pm$0.97 & \nodata & & & N & & & \\
C27  & 17:47:14.67, $-$28:37:49.51 & 0.03 & 2.05$\pm$0.19 & 2.16$\pm$0.19 & $-$2.10$\pm$0.39 & \nodata & & & N & & & \\
C28  & 17:46:28.11, $-$28:38:11.22 & 0.03 & 0.64$\pm$0.11 & 1.07$\pm$0.12 & $-$3.41$\pm$1.70 & \nodata & & & N & & & \\
C29  & 17:46:33.35, $-$28:38:29.52 & 0.04 & 0.98$\pm$0.14 & 1.52$\pm$0.15 & $-$4.31$\pm$1.50 & \nodata & & & N & & & \\
C30  & 17:46:06.18, $-$28:39:46.76 & 0.10 & 8.03$\pm$0.45 & 12.89$\pm$0.66&    0.13$\pm$0.01 & \nodata & & & N & & C(I12), JVLA2(R13)& \\
C31  & 17:46:43.00, $-$28:40:13.89 & 0.12 & 7.45$\pm$0.45 & 30.84$\pm$1.55& $-$0.95$\pm$0.04 & \nodata & & & N & & 2LC000.352-0.068(L08) & \\
C32  & 17:46:16.12, $-$28:40:15.36 & 0.09 & 1.50$\pm$0.12 & 4.07$\pm$0.22 & $-$0.56$\pm$0.13 & \nodata & & & N & & JVLA7(R13) & \\
C33  & 17:46:59.46, $-$28:41:09.55 & 0.04 & 0.85$\pm$0.09 & 1.29$\pm$0.10 & $-$0.93$\pm$0.44 & \nodata & & & N & & &  \\
C34  & 17:46:20.18, $-$28:41:28.29 & 0.02 & 0.59$\pm$0.11 & 0.76$\pm$0.11 & $-$2.87$\pm$1.72 & \nodata & & & N & & & \\
C35  & 17:46:39.62, $-$28:41:28.31 & 0.12 & 1.04$\pm$0.09 & 7.15$\pm$0.37 & $-$1.50$\pm$0.34 & \nodata & & & Y & & & \\
C36  & 17:46:04.17, $-$28:42:33.54 & 0.09 & 3.06$\pm$0.20 & 8.32$\pm$0.43 &    0.12$\pm$0.02 & \nodata & & & N & & B(I12), JVLA1(R13) & \\
C37  & 17:46:13.79, $-$28:43:43.84 & 0.05 & 0.81$\pm$0.11 & 1.88$\pm$0.14 & $-$0.41$\pm$0.17 & \nodata & & SSTGC618018? & Y & & JVLA6(R13) & UC? \\
C38  & 17:46:23.07, $-$28:45:56.20 & 0.06 & 0.99$\pm$0.13 & 2.76$\pm$0.18 & $-$3.01$\pm$1.01 & N & & & N & & & \\
C39  & 17:46:23.70, $-$28:48:22.00 & 0.05 & 0.79$\pm$0.13 & 2.57$\pm$0.18 & $-$2.33$\pm$0.73 & Y & & & N & & & \\
C40  & 17:46:21.00, $-$28:50:02.19 & 0.11 & 27.38$\pm$1.51& 64.18$\pm$3.22&    0.77$\pm$0.01 & N & & & Y & & N3(Y87) & \\
C41  & 17:45:50.43, $-$28:49:21.66 & 0.04 & 2.05$\pm$0.25 & 2.41$\pm$0.26 &    2.18$\pm$0.96 & Y & & & N & WR*(L05), X(M06,M09) & AR1(L05) & \\
C42  & 17:45:46.82, $-$28:49:54.38 & 0.06 & 2.18$\pm$0.36 & 5.76$\pm$0.44 &    0.31$\pm$0.12 & Y & & & N & & & UC? \\
C43  & 17:45:55.48, $-$28:50:10.81 & 0.07 & 1.22$\pm$0.20 & 4.63$\pm$0.30 &    0.16$\pm$0.08 & Y & & & N & Mi*(G01) & & \\
C44  & 17:45:43.88, $-$28:51:32.79 & 0.08 & 2.71$\pm$0.32 & 7.46$\pm$0.47 &    0.29$\pm$0.08 & Y & & & N & F*?(H19) & H13(L01) & \\
C45  & 17:45:43.75, $-$28:52:26.50 & 0.08 & 4.94$\pm$0.43 & 7.76$\pm$0.51 & $-$0.41$\pm$0.08 & Y & & & N & F*?(H19) & H12(L01) & \\
C46  & 17:45:38.83, $-$28:52:31.61 & 0.05 & 3.55$\pm$0.33 & 5.39$\pm$0.38 & $-$0.39$\pm$0.09 & Y & & & Y & F*?(H19) & H11(L01) & \\
C47  & 17:46:02.98, $-$28:52:44.33 & 0.14 & 16.57$\pm$0.93& 48.27$\pm$2.43& $-$0.29$\pm$0.01 & Y & & & N & PN(P16) & N1(Y87), 2LC000.098-0.051(L08) & \\
C48  & 17:45:58.32, $-$28:53:32.39 & 0.07 & 4.03$\pm$0.29 & 6.05$\pm$0.35 &    0.31$\pm$0.07 & N & & & N & X(Z18) & & \\
C49  & 17:46:09.93, $-$28:55:05.23 & 0.12 & 14.98$\pm$0.87& 39.66$\pm$2.01& $-$0.39$\pm$0.02 & N & & & N & & 2LC000.078-0.093(L08) & \\
C50  & 17:46:10.51, $-$28:55:49.02 & 0.07 & 7.21$\pm$0.46 & 9.99$\pm$0.56 &    0.72$\pm$0.09 & N & & & N & & & \\
C51  & 17:45:16.17, $-$29:03:15.04 & 0.02 & 1.10$\pm$0.16 & 1.09$\pm$0.16 &    1.25$\pm$1.03 & Y & & & N & WR*(M10), X(M09,Z18) & & \\
C52  & 17:45:24.04, $-$29:04:21.09 & 0.05 & 1.41$\pm$0.13 & 1.62$\pm$0.13 & $-$0.93$\pm$0.60 & N & & & N & & & \\
C53  & 17:45:11.82, $-$29:04:53.99 & 0.04 & 1.44$\pm$0.11 & 1.24$\pm$0.10 & $-$0.53$\pm$0.35 & N & & & N & & & \\
C54  & 17:44:58.63, $-$29:06:07.32 & 0.04 & 0.39$\pm$0.08 & 0.78$\pm$0.08 & $-$0.32$\pm$0.19 & Y & & & N & & & UC? \\
C55  & 17:45:17.81, $-$29:06:53.45 & UR   & 0.42$\pm$0.06 & 0.32$\pm$0.06 & $-$4.12$\pm$3.31 & N & & & N & X(M09) & & \\
C56  & 17:45:22.12, $-$29:10:59.18 & 0.16 & 4.04$\pm$0.24 & 15.28$\pm$0.77& $-$0.71$\pm$0.13 & Y & & SSTGC476516 & N & & & UC? \\
C57  & 17:45:40.82, $-$29:12:10.31 & UR   & 0.54$\pm$0.10 & 0.50$\pm$0.10 & $-$2.19$\pm$0.58 & \nodata & & & N & & & \\
C58  & 17:45:00.55, $-$29:10:07.70 & 0.04 & 0.58$\pm$0.08 & 1.15$\pm$0.10 & $-$1.86$\pm$0.58 & N & & & N & & & \\
C59  & 17:45:06.96, $-$29:10:36.21 & 0.04 & 0.66$\pm$0.13 & 1.69$\pm$0.15 & $-$2.35$\pm$0.70 & Y & & & N & & & \\
C60  & 17:45:01.08, $-$29:10:44.60 & 0.06 & 2.49$\pm$0.16 & 3.08$\pm$0.18 &    0.11$\pm$0.01 & N & & & N & X(M06,M09) & & \\
C61  & 17:44:54.84, $-$29:11:22.07 & 0.05 & 0.52$\pm$0.06 & 0.85$\pm$0.07 & $-$0.32$\pm$0.13 & N & & & N & & & \\
C62  & 17:45:01.85, $-$29:11:54.80 & 0.07 & 0.35$\pm$0.07 & 2.04$\pm$0.12 & $-$0.65$\pm$0.26 & N & & & N & & & \\
C63  & 17:45:08.96, $-$29:12:15.81 & 0.02 & 0.53$\pm$0.10 & 0.68$\pm$0.10 & $-$0.46$\pm$0.19 & Y & & SSTGC441299 & N & X(06) & & \\
C64  & 17:45:09.48, $-$29:12:23.61 & 0.05 & 0.79$\pm$0.10 & 1.61$\pm$0.12 & $-$1.95$\pm$0.51 & Y & & & N & & & \\
C65  & 17:45:10.08, $-$29:12:28.41 & 0.02 & 0.49$\pm$0.10 & 0.63$\pm$0.10 & $-$1.49$\pm$0.67 & Y & & & N & & & \\
C66  & 17:45:21.95, $-$29:13:47.88 & 0.14 & 4.36$\pm$0.24 & 19.92$\pm$1.00 &   0.40$\pm$0.01 & Y & & & N & & 2LC359.720-0.106(L08) & UC? \\
C67  & 17:45:32.57, $-$29:15:33.08 & 0.01 & 0.42$\pm$0.08 & 0.48$\pm$0.08 & $-$1.90$\pm$0.70 & \nodata & & & N & & & \\
C68  & 17:45:22.87, $-$29:15:50.27 & 0.07 & 2.52$\pm$0.14 & 2.94$\pm$0.15 &    0.41$\pm$0.04 & \nodata & & & N & & & \\
C69  & 17:44:40.13, $-$29:18:17.72 & 0.04 & 0.20$\pm$0.04 & 0.47$\pm$0.04 & $-$2.05$\pm$1.34 & \nodata & & & N & & & \\
C70  & 17:45:26.36, $-$29:18:21.44 & 0.03 & 0.47$\pm$0.05 & 0.48$\pm$0.05 & $-$0.60$\pm$0.31 & \nodata & & & N & & & \\
C71  & 17:45:31.39, $-$29:19:35.80 & 0.02 & 0.25$\pm$0.05 & 0.34$\pm$0.05 & $-$2.10$\pm$1.02 & \nodata & & & N & & & \\
C72  & 17:45:00.61, $-$29:19:38.60 & 0.04 & 0.33$\pm$0.05 & 0.66$\pm$0.06 & $-$2.99$\pm$1.75 & \nodata & & SSTGC419271 & N & & & \\
C73  & 17:45:32.36, $-$29:20:02.79 & 0.04 & 0.50$\pm$0.05 & 0.65$\pm$0.06 & $-$3.14$\pm$1.30 & \nodata & & & N & & & \\
C74  & 17:44:36.42, $-$29:20:25.17 & 0.08 & 1.54$\pm$0.12 & 2.67$\pm$0.16 & $-$2.34$\pm$0.33 & \nodata & G359.541$-$00.023 & & N & & 2LC359.540-0.023(L08) & \\
C75  & 17:44:43.58, $-$29:20:47.77 & 0.06 & 0.59$\pm$0.07 & 1.41$\pm$0.09 & $-$1.62$\pm$0.61 & \nodata & & SSTGC374813 & N & & & \\
C76  & 17:44:57.83, $-$29:21:24.49 & 0.01 & 0.16$\pm$0.03 & 0.20$\pm$0.03 & $-$2.06$\pm$1.71 & \nodata & & & N & & & \\
C77  & 17:44:39.63, $-$29:22:26.42 & 0.08 & 0.51$\pm$0.10 & 3.54$\pm$0.20 & $-$2.42$\pm$0.60 & \nodata & & & N & & & \\
C78  & 17:45:20.36, $-$29:22:33.18 & 0.04 & 0.16$\pm$0.03 & 0.36$\pm$0.03 & $-$2.03$\pm$1.61 & \nodata & & & N & & & \\
C79  & 17:45:18.57, $-$29:23:05.29 & 0.03 & 0.29$\pm$0.03 & 0.33$\pm$0.03 & $-$2.35$\pm$1.64 & \nodata & & & N & & & \\
C80  & 17:45:15.06, $-$29:23:27.51 & 0.01 & 0.14$\pm$0.03 & 0.16$\pm$0.03 & $-$0.43$\pm$0.38 & \nodata & & & N & & & \\
C81  & 17:45:19.45, $-$29:23:28.39 & UR   & 0.13$\pm$0.03 & 0.13$\pm$0.03 & $-$2.51$\pm$1.90 & \nodata & & & N & & & \\
C82  & 17:45:39.07, $-$29:23:30.00 & 0.04 & 0.46$\pm$0.05 & 0.60$\pm$0.05 & $-$0.67$\pm$0.31 & \nodata & & EGO g26 & Y & & & UC \\
C83  & 17:45:39.35, $-$29:23:30.90 & 0.03 & 0.39$\pm$0.05 & 0.48$\pm$0.05 & $-$1.89$\pm$0.98 & \nodata & & EGO g26 & Y & & & UC \\
C84  & 17:45:19.38, $-$29:23:35.59 & 0.05 & 0.13$\pm$0.03 & 0.47$\pm$0.03 & $-$2.35$\pm$1.79 & \nodata & & & N & & & \\
C85  & 17:45:11.83, $-$29:23:37.71 & 0.02 & 0.16$\pm$0.03 & 0.21$\pm$0.03 & $-$1.72$\pm$1.38 & \nodata & & & Y & & & \\
C86  & 17:45:37.24, $-$29:23:49.53 & 0.02 & 0.27$\pm$0.04 & 0.30$\pm$0.04 &    0.17$\pm$0.11 & \nodata & & & Y & X(M06) & & \\
C87  & 17:45:11.23, $-$29:24:16.11 & 0.05 & 0.29$\pm$0.04 & 0.57$\pm$0.04 & $-$3.27$\pm$2.27 & \nodata & & & Y & & & \\
C88  & 17:44:50.13, $-$29:24:19.63 & 0.05 & 0.44$\pm$0.05 & 0.70$\pm$0.06 & $-$1.99$\pm$1.05 & \nodata & & & N & & & \\
C89  & 17:44:53.25, $-$29:24:33.76 & 0.08 & 0.37$\pm$0.05 & 1.68$\pm$0.09 & $-$2.85$\pm$1.15 & \nodata & G359.513$-$00.111 & SSTGC400062 & N & & & UC? \\
C90  & 17:45:29.53, $-$29:24:43.02 & 0.04 & 0.69$\pm$0.05 & 0.68$\pm$0.05 & $-$1.41$\pm$0.44 & \nodata & & & N & & & \\
C91  & 17:45:14.47, $-$29:24:44.31 & 0.02 & 0.15$\pm$0.03 & 0.24$\pm$0.03 & $-$4.05$\pm$2.48 & \nodata & & & N & & & \\
C92  & 17:45:06.55, $-$29:24:56.91 & 0.08 & 0.20$\pm$0.03 & 1.07$\pm$0.06 & $-$1.84$\pm$1.19 & \nodata & & & N & & & \\
C93  & 17:45:15.22, $-$29:25:03.51 & 0.00 & 0.15$\pm$0.03 & 0.17$\pm$0.03 & $-$2.02$\pm$1.39 & \nodata & & & N & & & \\
C94  & 17:45:36.30, $-$29:25:22.24 & 0.06 & 0.30$\pm$0.05 & 1.21$\pm$0.08 & $-$4.07$\pm$1.18 & \nodata & & & N & & & \\
C95  & 17:44:31.89, $-$29:25:46.70 & 0.05 & 1.43$\pm$0.17 & 2.96$\pm$0.21 & $-$2.01$\pm$0.28 & \nodata & & & Y & & 2LC359.456-0.055(L08) & \\
C96  & 17:44:54.35, $-$29:25:58.07 & 0.10 & 0.37$\pm$0.05 & 2.31$\pm$0.12 & $-$2.14$\pm$0.91 & \nodata & & & Y & & & \\
C97  & 17:44:49.25, $-$29:26:14.83 & 0.03 & 0.25$\pm$0.05 & 0.41$\pm$0.05 & $-$1.46$\pm$1.17 & \nodata & & SSTGC390425 & Y & & & UC? \\
C98  & 17:45:14.84, $-$29:26:43.71 & 0.03 & 0.23$\pm$0.05 & 0.39$\pm$0.05 & $-$1.90$\pm$1.01 & \nodata & & & N & & & \\
C99  & 17:44:44.54, $-$29:27:02.18 & 0.04 & 3.74$\pm$0.22 & 3.35$\pm$0.19 &    0.69$\pm$0.04 & \nodata & & & Y & & & UC? \\
C100 & 17:45:20.44, $-$29:27:46.98 & 0.04 & 1.05$\pm$0.10 & 1.34$\pm$0.10 & $-$2.54$\pm$0.59 & \nodata & & & N & & & \\
C101 & 17:44:39.78, $-$29:27:50.12 & 0.07 & 0.97$\pm$0.17 & 4.88$\pm$0.29 & $-$1.69$\pm$0.23 & \nodata & & & Y & & & \\
C102 & 17:44:41.16, $-$29:27:54.94 & 0.03 & 1.08$\pm$0.14 & 1.41$\pm$0.15 & $-$0.35$\pm$0.07 & \nodata & & & Y & & H1(L19) & UC \\
C103 & 17:44:40.17, $-$29:28:14.72 & 0.07 & 4.37$\pm$0.30 & 6.16$\pm$0.36 & $-$0.53$\pm$0.03 & \nodata & & EGO g29 & Y & & 359.44$-$0.10A(F00), H3(L19)& UC \\
C104 & 17:44:40.58, $-$29:29:03.03 & 0.07 & 6.31$\pm$0.37 & 7.03$\pm$0.38 & $-$0.63$\pm$0.02 & \nodata & & & Y & & 359.44$-$0.10B(F00), 2LC359.425-0.111(L08) & \\
%%%%
\enddata
\tablenotetext{a}{Unresolved sources are marked by `UR'.}
\tablenotetext{b}{Intensities, fluxes, and spectral indices have been corrected for primary beam response.}
\tablenotetext{c}{The systematic uncertainty in spectral indices of 0.05 is not listed.}
\tablenotetext{d}{Counterparts that would argue against UC \hii{} regions, with references in parentheses. The object types generally follow the abbreviation convention of SIMBAD: X -- X-ray sources; Mi* -- Mira variable stars; F*? -- possible field stars; WR* -- Wolf-Rayet stars; PN -- planetary nebula.}
\tablenotetext{e}{Identifiers of previous detections of compact radio continuum emission using radio interferometers, with references in parentheses.}
\tablerefs{A11: \citet{an2011}. A14: \citet{anderson2014}. B11: \citet{battersby2011}. F00: \citet{forster2000}. G01: \citet{glass2001}. H19: \citet{hankins2019}. I12: \citet{immer2012b}. L01: \citet{lang2001}. L05: \citet{lang2005}. L19: \citet{lu2019a}. M10: \citet{mauerhan2010a}. M06: \citet{muno2006}. M09: \citet{muno2009}. P16: \citet{parker2016}. R13: \citet{rodriguez2013}. W10: \citet{wang2010}. Y87: \citet{yusefzadeh1987a}. Y09: \citet{yusefzadeh2009}. Z18: \citet{zhu2018}.}
\end{deluxetable*}
\end{longrotatetable}

%\clearpage

\begin{deluxetable*}{ccccccc}
\tablecaption{Properties of \meth{} masers.\label{tab:ch3ohmaser}}
\tablewidth{0pt}
\tablehead{
\colhead{ID} & R.A. \& Decl.  & $v_\text{peak}$\tablenotemark{a} & $F_\text{peak}$\tablenotemark{a,b} & $F_\text{int}$\tablenotemark{b} & Refs.\\
 & (J2000) & (\kms{}) & (Jy per channel) & (Jy\,\kms{}) & 
}
\startdata
M1    & 17:47:24.74, $-$28:21:43.16   & 68.17 & 21.22  & 50.82 & H95, C96 \\
M2   & 17:47:18.66, $-$28:22:10:17   & 71.68 &   0.06  & 0.40 & this work \\
M3   & 17:47:19.79, $-$28:22:12:64   & 73.44 &   0.05  & 0.24 & this work \\
M4   & 17:47:19.27, $-$28:22:14:50   & 73.44 &   0.77  & 4.79  & H95, C96 \\
M5   & 17:47:19.54, $-$28:22:31:98   & 60.79 &  0.08  & 0.11  & H95, C96 \\
M6   & 17:47:20.04, $-$28:22:40:98 & 58.34 &  8.45  & 6.01  & H95, C96 \\
M7   & 17:47:18.65, $-$28:22:54:28  & 70.63 & 37.64  & 33.88 & H95, C96 \\
M8  & 17:47:19.87, $-$28:23:01:04   & 55.17 &  0.35   & 0.37 & H95, C96 \\
M9   & 17:47:20.15, $-$28:23:06:02 & 66.76 &  0.16  & 0.19 & this work \\
M10 & 17:47:20.04, $-$28:23:12:46 & 60.44 & 4.55  & 2.66 & H95, C96 \\
M11 & 17:47:20.05, $-$28:23:46:42 & 50.61 & 5.02  & 10.05 & H95, C96 \\
M12 & 17:47:21.11, $-$28:24:17:95   & 48.15 & 17.71  & 16.02 & H95, C96 \\
M13 & 17:47:18.65, $-$28:24:24:63 & 49.55 & 56.06  & 102.11 & H95, C96 \\
M14 & 17:47:22.04, $-$28:24:42:33 & 50.96 & 3.46  & 3.34 & H95, C96 \\
M15 & 17:46:47.07, $-$28:32:06:86 & 27.90 & 2.00  & 4.80 & C09 \\
M16 & 17:46:21.41, $-$28:35:39:02 & 37.03 & 0.33  & 0.43 & C96 \\
M17 & 17:46:07.68, $-$28:45:20:48 & 49.33 & 2.94  & 5.31 & C96 \\
M18 & 17:45:35.00, $-$29:23:15:05 & 29.69 & 0.20  & 0.15 & this work \\
M19 & 17:45:39.86, $-$29:23:23:21 & 21.96 & 0.12  & 0.15 & this work \\
M20& 17:45:39.07, $-$29:23:29:96 & 19.50  & 41.23 & 64.62 & C96 \\
M21 &  17:44:41.31, $-$29:27:58:32 & $-$55.66 & 0.24  & 0.24 & this work \\
M22& 17:44:40.18, $-$29:28:12:28 & $-$56.72 & 0.98  & 1.91 & C96 \\
M23& 17:44:40.61, $-$29:28:15:28 & $-$46.88 & 63.92  & 61.13 & C96 \\
\enddata
\tablenotetext{a}{For masers with multiple velocity components along the line of sight, $V_\text{lsr}$ and flux of the strongest peak is listed.}
\tablenotetext{b}{Fluxes have been corrected for primary beam response.}
\tablerefs{C96: \citealt{caswell1996}. C09: \citealt{caswell2009}. H95: \citealt{houghton1995}.}
\end{deluxetable*}

\begin{deluxetable*}{ccccccc}
\tablecaption{Properties of \fmh{} masers.\label{tab:h2comaser}}
\tablewidth{0pt}
\tablehead{
\colhead{ID} & R.A. \& Decl.  & $v_\text{peak}$\tablenotemark{a} & $F_\text{peak}$\tablenotemark{a,b} & $F_\text{int}$\tablenotemark{b} & Refs. \\
 & (J2000) & (\kms{}) & (Jy per channel) & (Jy\,\kms{}) & 
}
\startdata
F1    & 17:47:19.86, $-$28:22:12:82 & 75.15 & 1.18\tablenotemark{c} & 0.91\tablenotemark{c} & M94 \\
F2   & 17:47:20.14, $-$28:23:04:23 & 50.41 & 0.09\tablenotemark{c} & 0.66\tablenotemark{c} & M94 \\
F3   & 17:47:20.05, $-$28:23:46:58 & 49.93 & 0.21 & 0.70 & M94 \\
F4   & 17:47:19.58, $-$28:23:49:64 & 48.96 & 0.18 & 0.38 & M94 \\
F5   & 17:47:18.65, $-$28:24:24:57 & 48.96 & 0.29\tablenotemark{c} & 0.64\tablenotemark{c} & M94\\
F6   & 17:46:21.40, $-$28:35:39:09 & 36.96 & 0.88 & 1.54 & G15 \\
F7   & 17:44:40.18, $-$29:28:11:96 & $-$43.03 & 0.09 & 0.13 & this work \\
F8  & 17:44:40.60, $-$29:28:15:10 & $-$54.19 & 0.18 & 0.44 & this work \\
\enddata
\tablenotetext{a}{For masers with multiple velocity components along the line of sight, $V_\text{lsr}$ and flux of the strongest peak is listed.}
\tablenotetext{b}{Fluxes have been corrected for primary beam response.}
\tablenotetext{c}{Significant absorption is superimposed with maser emission (\autoref{fig:maserspec}), therefore reported fluxes are lower limits.}
\tablerefs{G15: \citealt{ginsburg2015}. M94: \citealt{mehringer1994}.}
\end{deluxetable*}

\begin{deluxetable*}{c@{\hspace{1em}}c@{\hspace{1em}}c@{\hspace{1em}}c@{\hspace{1em}}c@{\hspace{1em}}c}
\tablecaption{Summary of High-mass Star Formation Indicators.\label{tab:sfsummary}}
\tablewidth{0pt}
\tablehead{
\colhead{Regions} & No.\ of UC/HC \hii{} Regions & Refs.\ for UC/HC \hii{} Regions & No.\ of \Clt{} \meth{} Masers & No.\ of Unique Indicators\tablenotemark{a} & SFR\tablenotemark{b} (10$^{-3} $\msolpyr{})
}
\startdata
Sgr~B2                     &  41 & G95, D15         & 14  & 50  & 21.5 \\
Dust Ridge cloud e &  1  & L19                    & 1    & 1    & 0.4  \\
Dust Ridge cloud c &  1  & this work         & 1    & 1    & 0.4 \\
Brickette                   & 0  & \nodata & 1    & 1    & 0.4 \\
\Cfi{}                         & 1   & E83, M11, L19 & 0    & 1    & 0.4  \\
\Ctw{}                       & 1   & L19                    & 0    & 1    & 0.4  \\
Sgr~C                        & 2   & F00, L19, this work  & 3    & 3  & 1.3 \\
\hline
Total                          & 47 & \nodata            & 20 & 58 & 24.9
\enddata
\tablenotetext{a}{\Clt{} \meth{} masers spatially associated with known UC/HC \hii{} regions are excluded. In Sgr B2: M6 -- source Z10.24 \citep{gaume1995}. M7 -- source Y \citep{gaume1995}. M8 -- source B \citep{gaume1995}. M9 -- sources F10.33/F10.35/F10.37/F10.39 \citep{gaume1995}. M10 -- source D \citep{gaume1995}. In Dust Ridge cloud e: M15 -- source H1 \citep{lu2019a}. In Dust Ridge cloud c: M16 -- C23 (this work). In Sgr C: M21 -- C102 (this work). M22 -- C103 (this work).}
\tablenotetext{b}{We assume each unique high-mass star formation indicator corresponds to a high-mass protostar of $>$10~\msol{}, thus representing a total stellar mass of 129~\msol{} \citep[see Appendix D of][]{lu2019a}. The SFR is derived by dividing the total stellar mass in the considered region by the characteristic time scale of 0.3~Myr.}
\tablerefs{D15: \citealt{depree2015}. E83: \citealt{ekers1983}. F00: \citealt{forster2000}. G95: \citealt{gaume1995}. L19: \citealt{lu2019a}. M11: \citealt{mills2011}.}
\end{deluxetable*}

\end{CJK}
\end{document}